\documentclass{aa}  

\usepackage{graphicx}
\usepackage{txfonts}
\usepackage{natbib} 
\bibpunct{(}{)}{;}{a}{}{,} % to follow the A&A style

\usepackage{verbatim}
\usepackage{url}
\usepackage{epstopdf}

\begin{document}

   \title{Uncertainties in asteroseismic grid-based estimates of stellar ages}

   \subtitle{SCEPtER: Stellar CharactEristics Pisa
Estimation gRid}

   \author{G. Valle \inst{1,2,3}, M. Dell'Omodarme \inst{3}, P.G. Prada Moroni
     \inst{2,3}, S. Degl'Innocenti \inst{2,3} 
          }

   \authorrunning{Valle, G. et al.}

   \institute{
INAF - Osservatorio Astronomico di Collurania, Via Maggini, I-64100, Teramo, Italy 
\and
 INFN,
 Sezione di Pisa, Largo Pontecorvo 3, I-56127, Pisa, Italy
\and
Dipartimento di Fisica ``Enrico Fermi'',
Universit\`a di Pisa, Largo Pontecorvo 3, I-56127, Pisa, Italy
 }

   \offprints{G. Valle, valle@df.unipi.it}

   \date{Received 26/07/2014; accepted 08/12/2014}

  \abstract
{Stellar age determination by means of grid-based techniques 
adopting asteroseismic constraints
is nowadays a
  well established method. However some theoretical aspects of the systematic
  and statistical errors affecting these age estimates have still to be
  investigated.} 
   {  
We study the impact on stellar age determination of the uncertainty in
 the radiative opacity, in the initial helium abundance, in the
 mixing-length value, in the convective core
 overshooting, and in the microscopic
 diffusion efficiency adopted in stellar model computations.
}
{  
We extended our SCEPtER grid (Valle et al. 2014) to include stars
with mass in the range 
[0.8 - 1.6] $M_{\sun}$ and evolutionary 
stages from the zero-age main sequence to the central hydrogen depletion. 
For the age estimation we adopted  the same maximum likelihood technique
described in our previous
work. To quantify the 
systematic errors arising from the current uncertainty in model computations, many
synthetic grids of stellar models with perturbed input were adopted.
}
  {
We found that the current typical
uncertainty in the observations accounts for $1 \sigma$ statistical relative error in
age determination which in mean ranges from about $-35\%$ to $+42\%$, depending on the mass. 
However, due to the strong dependence on the evolutionary phase, the age relative error
 can be higher than 120\% for stars near the zero-age
main-sequence, while it is typically of the order of 20\% or lower in the advanced main-sequence phase.  
The systematic bias on age determination due to a variation of  $\pm 1$ in the
helium-to-metal enrichment ratio
$\Delta Y/\Delta Z$ is about one-forth of the statistical error in the first
30\% of the evolution while it is negligible for more evolved stages.
The maximum bias due to the presence of the convective core overshooting is of $-7\%$ and $-13\%$
for mild and strong overshooting scenarios. 
For all the examined models the impact of  a variation of $\pm 5\%$ in the
radiative opacity was found to be negligible.    
The most important source of bias are the uncertainty in the mixing-length value
$\alpha_{\rm ml}$ and the neglect of microscopic diffusion. Each of these
effects accounts for a bias which is nearly equal to the random error
uncertainty. 
Comparison of the results of our technique with other
grid techniques on a set of common stars showed a general agreement. However,
the adoption of a different grid can account for a variation in the mean
estimated age up to 1 Gyr.  
 }
{}

   \keywords{
Asteroseismology --
methods: statistical --
stars: evolution --
stars: oscillations --
stars: low-mass --
stars: fundamental parameters 
}

   \maketitle

\section{Introduction}\label{sec:intro}

The determination of stellar ages can not be obtained by direct
measurements; therefore, several techniques -- with different scope
of applicability -- 
have been developed to obtain sensible age estimates \citep[see][for a
  review]{Soderblom2010}. 
Traditionally the age estimate for a main-sequence single star involves
either the use of a rotation-mass-age relationship
or the comparison of computed isochrones
to observed parameters, which are classically magnitudes, colours,
 and metallicity. It is however well known that the 
precision of these estimates is unsatisfactory \citep[see
  e.g.][and references therein]{Lebreton2009, Lebreton2013, Epstein2014}.

The availability of high quality data from asteroseismology satellite missions,
such as CoRoT \citep[see e.g.][]{Appourchaux2008,Michel2008,Baglin2009} and
{\it Kepler} \citep[see e.g.][]{Borucki2010, Gilliland2010}, has offered a
great contribution to stellar age estimation. These data, combined with the
traditional ones, 
increase the constraints on the theoretical models allowing a significant
improvement.

It has been recently shown that the
single star modelling in presence of individual frequencies of the oscillation
spectrum can provide age estimates precise better than 15\%, with a typical
factor of  
two improvement over grid based estimates adopting as seismic observables 
the average large frequency spacing $\Delta \nu$ and
the frequency of maximum oscillation power $\nu_{\rm max}$
\citep{Lebreton2013, SilvaAguirre2013, Metcalfe2014}. However these techniques
are computationally intensive and are restricted only to stars for which high
signal-to-noise 
data are available.   
The alternative approach based on grid techniques adopting $\Delta \nu$ and
$\nu_{\rm max}$ \citep[see 
  e.g.][]{Stello2009,Basu2010,Quirion2010,Gai2011,Huber2013} 
has been recognised to allow a fast and automated way to obtain stellar
ages from data. 
Although several studies have been  devoted to quantify 
the uncertainty affecting these grid-based estimates \citep[see e.g.][]{Gai2011,
  Lebreton2013} 
 a comprehensive examination of the various bias sources and
statistical uncertainties is still lacking. 
In this paper we continue the work started in \citet{scepter1} (hereafter
V14) -- focused on
mass and radius estimates --
 by exploring the
systematic biases on grid-based age estimates due to the uncertainties  
on the main error sources in theoretical predictions: the radiative opacity,
the microscopic diffusion 
efficiency,  the mixing-length
parameter value, the initial helium abundance-metallicity
relationship, and convective core overshooting extension.
We restrict our analysis to central hydrogen-burning stars with
mass in the range [0.8 - 1.6] $M_{\sun}$. 

The structure of the paper is the following. In Sect.~\ref{sec:method} we 
discuss the method and the grids used in the estimation process. 
The main
results are presented in Sects.~\ref{sec:results} and \ref{sec:errorprop}.
Sect.~\ref{sec:comparison} 
presents a comparison of the estimates obtained with our grids with those
by other grid-based techniques. 
Some
concluding remarks can be found in Sect.~\ref{sec:conclusions}.

\section{Grid-based recovery technique}\label{sec:method}

We adopted the SCEPtER scheme\footnote{An R library providing the estimation code and grid is available at CRAN: \url{http://CRAN.R-project.org/package=SCEPtER}.}
 described in V14  and derived from
 \citet{Basu2012}.  
For reader's convenience we summarise the basic aspects of the procedure.
We let $\cal
 S$ be a star for which the following vector of observed quantities
is available: $q^{\cal S} \equiv \{T_{\rm eff, \cal S}, {\rm [Fe/H]}_{\cal S},
\Delta \nu_{\cal S}, \nu_{\rm max, \cal S}\}$. Then we let $\sigma = \{\sigma(T_{\rm
  eff, \cal S}), \sigma({\rm [Fe/H]}_{\cal S}), \sigma(\Delta \nu_{\cal S}),
\sigma(\nu_{\rm max, \cal S})\}$ be the nominal uncertainty in the observed
quantities. For each point $j$ on the estimation grid of stellar models, 
we define $q^{j} \equiv \{T_{{\rm eff}, j}, {\rm [Fe/H]}_{j}, \Delta \nu_{j},
\nu_{{\rm max}, j}\}$. 
Let $ {\cal L}_j $ be the likelihood function defined as
\begin{equation}
{\cal L}_j = \left( \prod_{i=1}^4 \frac{1}{\sqrt{2 \pi} \sigma_i} \right)
\times \exp \left( -\frac{\chi^2}{2} \right)
\label{eq:lik}
\end{equation}
where
\begin{equation}
\chi^2 = \sum_{i=1}^4 \left( \frac{q_i^{\cal S} - q_i^j}{\sigma_i} \right)^2.
\end{equation}

The likelihood function is evaluated for each grid point within $3 \sigma$ of
all the variables from $\cal S$; let ${\cal L}_{\rm max}$ be the maximum value
obtained in this step. The estimated stellar mass, radius,
and age are obtained
by averaging the corresponding quantity of all the models with likelihood
greater than $0.95 \times {\cal L}_{\rm max}$.
Informative priors can be inserted as a multiplicative factor
in Eq.~(\ref{eq:lik}), as a weight attached to the
grid points.

The technique can also be employed to construct a Monte Carlo confidence
interval for mass, radius and age estimates. 
 To this purpose a synthetic sample of $n$ stars is
generated, following a multivariate normal distribution with vector of mean
$q^{\cal S}$ and covariance matrix $\Sigma = {\rm diag}(\sigma)$. A value of
$n = 10\,000$ is usually adopted since it provides a fair balance between
computation time and the accuracy of the results.  The medians of the $n$
objects mass, radius and age are taken as the best estimate of the true values;
the 
16th and 84th quantiles of the $n$ values are adopted as a $1 \sigma$
confidence interval.

\subsection{Standard stellar models grid}

The standard estimation grid of stellar models is obtained using FRANEC
stellar evolution code \citep{scilla2008}, in the same
configuration as was adopted to compute the Pisa Stellar
Evolution Data Base\footnote{\url{http://astro.df.unipi.it/stellar-models/}} 
for low-mass stars \citep{database2012, stellar}.

The grid consists of 141680 points (110 points for 1288 evolutionary tracks),
corresponding to evolutionary stages from the ZAMS to central hydrogen
depletion. Models are computed for masses in the range [0.80 - 1.60]
$M_{\sun}$ with a step of 0.01 $M_{\sun}$. The initial metallicity [Fe/H]
is assumed in the range [$-0.55$ - 0.55] with a step of 0.05 dex. The solar scaled heavy-element mixture by
\citet{AGSS09} is adopted.  The initial helium abundance is obtained using the
linear relation $Y = Y_p+\frac{\Delta Y}{\Delta Z} Z$
adopting a primordial $^4$He abundance value $Y_p = 0.2485$ from WMAP
\citep{cyburt04,steigman06,peimbert07a,peimbert07b}, and assuming $\Delta
Y/\Delta Z = 2$ \citep{pagel98,jimenez03,gennaro10}. The models are computed
assuming our solar-scaled mixing-length parameter $\alpha_{\rm
  ml} = 1.74$. Convective core overshooting is
not taken into account.
Atomic diffusion is included adopting the coefficients given by
  \citet{thoul94} for gravitational settling and thermal diffusion. 
To prevent the surface helium and metals depletion for stars
without a convective envelope, a diffusion inhibition mechanism similar to that
discussed in \citet{Chaboyer2001} is considered.
For the outermost 1\% in mass of the star the diffusion velocities are
multiplied by a suppression parabolic factor which takes value 1 at the 99\% in mass of the
structure and 0 at the base of the atmosphere.

Further details on the input adopted in the
computations are available in \citet{scepter1, cefeidi}.

As in V14, the average large frequency spacing $\Delta \nu$ and
the frequency of maximum 
oscillation power $\nu_{\rm max}$ are obtained using a simple scaling from
the solar values \citep{Ulrich1986, Kjeldsen1995}: 
\begin{eqnarray}\label{eq:dni}
\frac{\Delta \nu}{\Delta \nu_{\sun}} & = &
\sqrt{\frac{M/M_{\sun}}{(R/R_{\sun})^3}} \quad ,\\  \frac{\nu_{\rm
    max}}{\nu_{\rm max, \sun}} & = & \frac{{M/M_{\sun}}}{ (R/R_{\sun})^2
  \sqrt{ T_{\rm eff}/T_{\rm eff, \sun}} }. \label{eq:nimax}
\end{eqnarray}

\section{Age estimates: grid technique internal accuracy}\label{sec:results}

The age recovery procedure was first tested on a synthetic dataset obtained
by sampling $N = 50\,000$ artificial stars from the same standard estimation
grid of stellar models used in the recovery procedure itself and adding to each of them
a Gaussian noise in all the observed quantities. As in V14, we adopted the same standard
deviation values suggested by \citet{Gai2011}: i.e. 2.5\% in $\Delta \nu$, 5\% in 
$\nu_{\rm max}$, 100 K in $T_{\rm eff}$, and 0.1 dex in [Fe/H].

\begin{figure*}
\centering
\includegraphics[height=17cm,angle=-90]{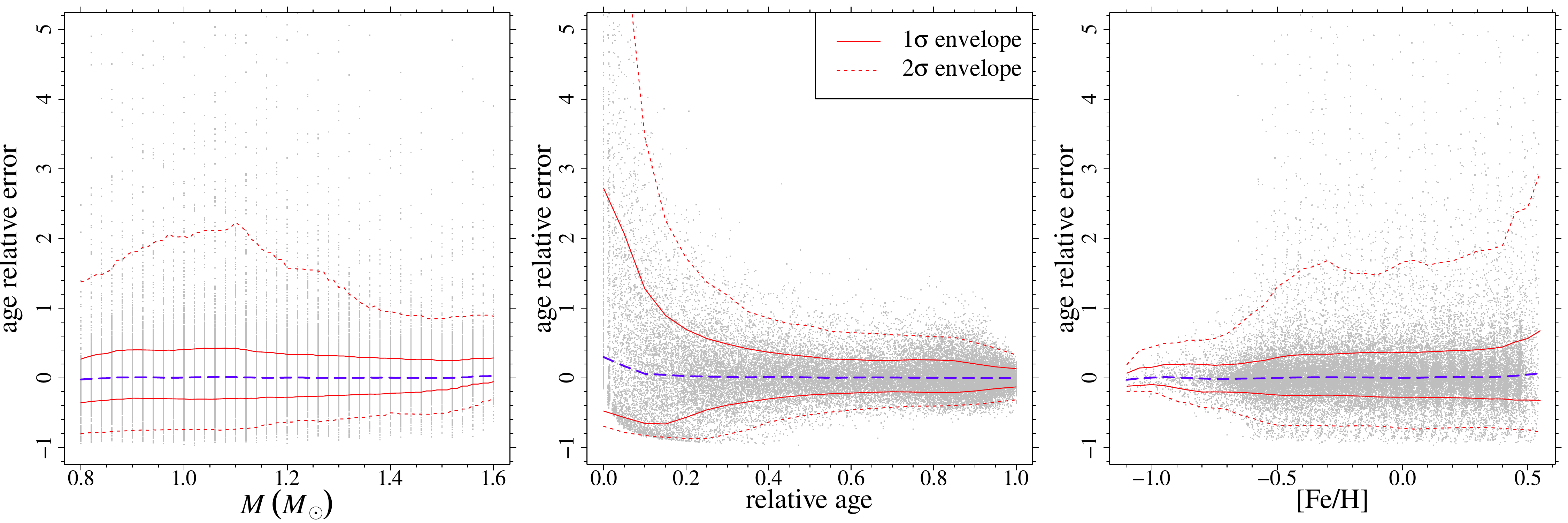}
\caption{Age estimate relative errors as a function of the true mass (left panel), 
  relative age (central panel), and metallicity [Fe/H] (right panel) of the
  star. The blue long dashed lines mark the error medians. The red solid line is the $1 \sigma$ error
  envelope, while the red dashed one marks the position of the $2 \sigma$
  envelope (see text). A positive relative error indicates that the reconstructed age of the star 
  is overestimated with respect to the true one. }
\label{fig:points}
\end{figure*}

For each synthetic star the relative error on the reconstructed age 
was computed. A positive relative error indicates that the age of the star is
overestimated by the recovery procedure. 
Figure~\ref{fig:points} shows the trend of the age relative errors versus
the 
true mass of the 
star, its relative age -- conventionally set to
0 at the ZAMS position and defined as the ratio between
the current age of the star and its age at 
the central hydrogen depletion -- and its metallicity [Fe/H]\footnote{This is the [Fe/H] value 
currently present on the stellar surface, not the initial one because of microscopic diffusion.}. 
The figure also shows the relative error envelopes obtained by evaluating
the 16th and 84th quantiles ($1 \sigma$) and 2.5th and 97.5th quantiles ($2
\sigma$)  
 of the age relative error over a moving window\footnote{The
half-width of the 
window is typically 1/12-1/16 of the range spanned by the independent
variable. This choice allows to maintain 
the mean relative error on the $1 \sigma$ envelope due to Monte Carlo sampling at a level of
about 5\%, without introducing too much smoothing.}. The position of the 1$\sigma$ 
envelope and of the median of the age relative error  
in dependence on the true mass of the star and on its relative age are
reported in Tab.~\ref{tab:global} and \ref{tab:global-pcage}, in the 
section labelled "standard".

As expected and already reported in literature \citep[see e.g.][]{Gai2011,
  Chaplin2014}, age determinations are much less constrained than mass and
radius ones. Moreover, the distribution of relative errors on age estimates is 
typically asymmetric and present a long tail toward age overestimate. This is due to the 
presence of the hard boundary at -1.0 since age estimates can not be negative.

Both the plot of the age relative errors versus the mass of the star and 
Tab.~\ref{tab:global} show that the overall $1 \sigma$ envelope ranges from about 
-35\% to +42\%. Moreover, it is evident the
presence of an "edge effect" similar to that extensively discussed in
V14. At the lower edge of the grid ($M$ = 0.80 $M_{\sun}$) the
age of the stars are biased toward low values, while the opposite occurs at
the upper edge ($M$ = 1.60 $M_{\sun}$). This trend is easily understood
considering that a star with $M$ = 0.80 $M_{\sun}$ can be confused in the
recovery with a slightly more massive model, which evolves faster and thus has a lower age, 
while no models at lower mass, and hence older, exist in the grid.  

The central panel in Fig.~\ref{fig:points} and Tab.~\ref{tab:global-pcage} show the strong dependence 
of the age relative error on the evolutionary phase: the closer the star is to the ZAMS and the
larger the uncertainty. The age relative error can be higher than 120\%  near the ZAMS, 
while it is typically of the order of 20\% or lower in the advanced main-sequence phase.
The high value of the upper envelope border at low
relative age is not surprising since an error on age estimates has a great impact here
since stellar ages are low. The envelope is highly asymmetric since, at low
relative ages, the grid edge 
limits the possibility to age underestimation thus resulting in biased
  estimates.  In other words, 
approaching the ZAMS, grid-based age estimates not only become considerably 
more uncertain but also biased toward older ages. 
The increase of the relative age leads to a shrink of the age relative error
envelope. 
A small envelope inflation is present around the relative age 0.8, which is
due to the presence of a convective core for the more massive models (see
Sect.~\ref{sec:ov});    
at relative age higher than about 0.85 the envelope shows a shrink,
due to the fact that age estimations are intrinsically easier in
rapid evolving phases \citep[see e.g. the results in][]{Gai2011, Chaplin2014}.

The trend versus [Fe/H] originates from the trend in relative age and from
edge effects. The
envelope of the  
relative error is narrower for [Fe/H] lower than about $-0.8$ dex, since for
these values only evolved models (relative age of about 0.8) are present. These
models reach such a low  
surface metallicity due to microscopic diffusion (we recall that
the lowest initial metallicity in the grid is [Fe/H] = $-0.55$ dex), which
takes long time  
scales to produce observable effects.
On the contrary, at the upper metallicity edge there must be only
models young  
enough for diffusion to be inefficient, i.e. models at very early evolutionary
stages,  
and consequently age estimates get less precise leading to a larger envelope. 

The medians ($q_{50}$) in the tables clearly
show the presence of the edge effect distortions discussed above.   

\begin{table*}[ht]
\centering
\caption{SCEPtER median ($q_{50}$) and $1 \sigma$ envelope boundaries
  ($q_{16}$ and 
  $q_{84}$) for age relative error as a 
  function of the mass of the star. Values
are espressed as percent.} 
\label{tab:global}
\begin{tabular}{lccccccccc}
  \hline\hline
 & \multicolumn{9}{c}{Mass ($M_{\sun}$)}\\
 & 0.80 & 0.90 & 1.00 & 1.10 & 1.20 & 1.30 & 1.40 & 1.50 & 1.60\\
\hline
  \multicolumn{10}{c}{standard}\\
  $q_{16}$ & -35.4 & -29.2 & -30.2 & -29.4 & -27.7 & -24.9 & -22.1 & -17.3 & -5.5 \\ 
  $q_{50}$ & -2.3 & 0.8 & 0.5 & 1.1 & 0.4 & -0.2 & 0.3 & -0.1 & 3.0 \\ 
  $q_{84}$ & 27.0 & 40.1 & 40.6 & 42.6 & 33.7 & 31.4 & 28.0 & 25.1 & 28.6 \\ 
\hline
  \multicolumn{10}{c}{$\sigma(T_{\rm eff})$ = 50 K}\\
  $q_{16}$ & -26.6 & -20.5 & -21.2 & -21.6 & -23.2 & -22.6 & -19.8 & -15.7 & -5.3 \\ 
  $q_{50}$ & -1.6 & 0.6 & 0.6 & 0.7 & -0.1 & 0.0 & 0.2 & 0.0 & 2.4 \\ 
  $q_{84}$ & 23.2 & 28.4 & 27.2 & 27.8 & 24.9 & 24.5 & 24.1 & 22.5 & 26.6 \\ 
\hline
  \multicolumn{10}{c}{$\sigma({\rm[Fe/H]})$ = 0.05 dex}\\
  $q_{16}$ & -32.5 & -27.6 & -28.2 & -27.2 & -24.1 & -21.4 & -18.3 & -13.8 & -5.1 \\ 
  $q_{50}$ & -1.1 & 0.5 & 0.3 & 1.2 & 0.3 & 0.1 & 0.0 & -0.0 & 2.4 \\ 
  $q_{84}$ & 26.7 & 35.8 & 35.8 & 38.1 & 28.3 & 26.0 & 22.6 & 19.3 & 21.2 \\ 
\hline
  \multicolumn{10}{c}{$\sigma(\Delta \nu, \nu_{\rm max})$ = 1\%, 2.5\%}\\
  $q_{16}$ & -30.7 & -26.4 & -27.2 & -26.2 & -21.7 & -19.5 & -16.4 & -13.4 & -4.6 \\ 
  $q_{50}$ & -1.0 & 0.2 & 0.0 & 0.8 & 0.0 & 0.0 & 0.0 & 0.0 & 1.9 \\ 
  $q_{84}$ & 24.2 & 34.0 & 34.4 & 36.9 & 27.2 & 23.4 & 20.7 & 17.2 & 17.6 \\ 
\hline
  \multicolumn{10}{c}{weighted}\\
  $q_{16}$ & -29.9 & -25.3 & -26.1 & -28.9 & -29.9 & -30.5 & -30.2 & -24.2 & -9.0 \\ 
  $q_{50}$ & 0.5 & 3.8 & 3.6 & 2.9 & -0.2 & -2.1 & -3.4 & -4.3 & 1.3 \\ 
  $q_{84}$ & 31.9 & 46.8 & 48.3 & 50.2 & 35.7 & 31.2 & 25.8 & 18.4 & 20.9 \\ 
\hline
  \multicolumn{10}{c}{$\Delta Y/\Delta Z$ = 1}\\
  $q_{16}$ & -37.1 & -25.1 & -24.1 & -22.8 & -25.7 & -24.6 & -23.9 & -19.5 & -11.5 \\ 
  $q_{50}$ & -5.8 & 5.6 & 4.7 & 3.9 & 1.6 & 0.4 & 0.6 & 0.3 & 1.5 \\ 
  $q_{84}$ & 23.4 & 53.0 & 49.8 & 49.7 & 41.7 & 34.7 & 31.1 & 29.5 & 30.6 \\ 
\hline
  \multicolumn{10}{c}{$\Delta Y/\Delta Z$ = 3}\\
  $q_{16}$ & -44.0 & -39.3 & -39.5 & -34.4 & -28.3 & -26.6 & -21.8 & -13.3 & -0.0 \\ 
  $q_{50}$ & -6.3 & -4.7 & -2.0 & -0.7 & -0.2 & -0.4 & 0.1 & 2.1 & 10.7 \\ 
  $q_{84}$ & 23.0 & 29.6 & 33.7 & 33.9 & 27.9 & 27.2 & 25.8 & 24.8 & 36.0 \\ 
\hline
  \multicolumn{10}{c}{$\alpha_{\rm ml}$ = 1.50}\\
  $q_{16}$ & -3.8 & 0.5 & -0.2 & -4.8 & -8.8 & -13.3 & -15.4 & -13.5 & -4.7 \\ 
  $q_{50}$ & 21.1 & 33.8 & 35.1 & 32.4 & 19.2 & 10.6 & 5.9 & 3.5 & 4.9 \\ 
  $q_{84}$ & 69.9 & 106.1 & 123.9 & 121.1 & 94.7 & 63.7 & 47.7 & 38.6 & 33.7 \\ 
\hline
  \multicolumn{10}{c}{$\alpha_{\rm ml}$ = 1.98}\\
  $q_{16}$ & -66.7 & -61.0 & -65.9 & -59.9 & -51.7 & -40.8 & -33.2 & -22.3 & -7.8 \\ 
  $q_{50}$ & -27.2 & -23.6 & -25.5 & -20.8 & -15.5 & -10.7 & -6.4 & -2.7 & 1.8 \\ 
  $q_{84}$ & 1.8 & 5.6 & 4.1 & 6.8 & 9.9 & 13.0 & 16.1 & 20.5 & 22.7 \\ 
\hline
  \multicolumn{10}{c}{no diffusion}\\
  $q_{16}$ & -4.6 & 0.8 & -1.1 & -1.1 & -2.1 & -6.6 & -9.2 & -10.1 & -1.7 \\ 
  $q_{50}$ & 22.9 & 38.9 & 37.4 & 38.2 & 31.5 & 22.9 & 16.5 & 11.5 & 11.1 \\ 
  $q_{84}$ & 65.0 & 100.6 & 104.6 & 106.6 & 90.9 & 77.7 & 60.8 & 50.1 & 47.7 \\ 
\hline
  \multicolumn{10}{c}{standard, restricted to $M$ > 1.10 $M_{\sun}$}\\
  $q_{16}$ &  &  &  &  & -27.4 & -25.8 & -23.8 & -17.4 & -7.6 \\ 
  $q_{50}$ &  &  &  &  & -0.2 & -0.4 & 0.1 & 0.5 & 2.8 \\ 
  $q_{84}$ &  &  &  &  & 29.3 & 30.5 & 28.1 & 26.6 & 26.6 \\ 
\hline
  \multicolumn{10}{c}{overshooting $\beta$ = 0.2}\\
  $q_{16}$ &  &  &  &  & -22.8 & -23.2 & -27.5 & -25.3 & -17.0 \\ 
  $q_{50}$ &  &  &  &  & -0.8 & -3.0 & -5.4 & -7.1 & -5.0 \\ 
  $q_{84}$ &  &  &  &  & 25.6 & 29.4 & 24.1 & 21.9 & 21.6 \\ 
\hline
  \multicolumn{10}{c}{overshooting $\beta$ = 0.4}\\
  $q_{16}$ &  &  &  &  & -39.0 & -37.0 & -34.0 & -31.0 & -23.0 \\ 
  $q_{50}$ &  &  &  &  & -13.5 & -11.5 & -11.8 & -13.4 & -10.1 \\ 
  $q_{84}$ &  &  &  &  & 13.7 & 18.7 & 15.3 & 11.6 & 10.5 \\ 
\hline
\end{tabular}
\tablefoot{Typical Monte Carlo relative uncertainty on $q_{16}$ and $q_{84}$ is about
  5\%, while the absolute uncertainty on $q_{50}$ is about 0.5\%.} 
\end{table*}

\begin{table*}[ht]
\centering
\caption{SCEPtER median ($q_{50}$) and $1 \sigma$ envelope boundaries
  ($q_{16}$ and 
  $q_{84}$) for age relative error as a
  function of the relative age of the star. Values
are expressed as percent.} 
\label{tab:global-pcage}
\begin{tabular}{lcccccccccc}
  \hline\hline
 & \multicolumn{10}{c}{relative age}\\
 & 0.1 & 0.2 & 0.3 & 0.4 & 0.5 & 0.6 & 0.7 & 0.8 & 0.9 & 1.0\\
\hline
  \multicolumn{11}{c}{standard}\\
  $q_{16}$ & -65.3 & -56.7 & -39.1 & -30.2 & -24.5 & -21.8 & -19.9 & -21.2 & -18.8 & -13.0 \\ 
  $q_{50}$ & 5.9 & 2.5 & 1.2 & 1.3 & 0.7 & 0.5 & 0.6 & 0.2 & -0.2 & -0.4 \\ 
  $q_{84}$ & 127.8 & 69.6 & 48.5 & 36.8 & 30.4 & 26.5 & 24.8 & 25.7 & 20.1 & 13.2 \\ 
\hline
  \multicolumn{11}{c}{$\sigma(T_{\rm eff})$ = 50 K}\\
  $q_{16}$ & -62.2 & -46.3 & -31.0 & -23.7 & -18.8 & -16.6 & -14.9 & -15.8 & -13.9 & -9.6 \\ 
  $q_{50}$ & 4.7 & 2.7 & 2.0 & 1.0 & 0.2 & 0.2 & 0.1 & 0.0 & -0.0 & -0.3 \\ 
  $q_{84}$ & 100.1 & 55.9 & 38.1 & 28.6 & 21.7 & 18.3 & 17.4 & 18.9 & 14.7 & 9.8 \\ 
\hline
  \multicolumn{11}{c}{$\sigma({\rm[Fe/H]})$ = 0.05 dex}\\
  $q_{16}$ & -62.5 & -48.4 & -33.7 & -25.9 & -22.2 & -19.3 & -17.8 & -19.2 & -17.1 & -11.8 \\ 
  $q_{50}$ & 3.8 & 1.3 & 0.7 & 1.2 & 0.9 & 0.6 & 0.6 & 0.4 & 0.0 & -0.3 \\ 
  $q_{84}$ & 107.2 & 58.4 & 40.3 & 30.6 & 25.4 & 22.8 & 21.7 & 23.9 & 18.2 & 11.5 \\ 
\hline
  \multicolumn{11}{c}{$\sigma(\Delta \nu, \nu_{\rm max})$ = 1\%, 2.5\%}\\
  $q_{16}$ & -58.3 & -43.8 & -32.0 & -25.0 & -21.2 & -18.9 & -17.1 & -18.2 & -16.2 & -11.6 \\ 
  $q_{50}$ & 3.9 & 0.3 & 0.0 & 0.5 & 0.0 & 0.0 & 0.0 & 0.0 & 0.0 & 0.0 \\ 
  $q_{84}$ & 95.5 & 52.6 & 37.0 & 29.4 & 25.3 & 22.0 & 21.4 & 22.4 & 17.1 & 10.9 \\ 
\hline
  \multicolumn{11}{c}{weighted}\\
  $q_{16}$ & -61.4 & -48.1 & -34.2 & -25.6 & -21.1 & -18.7 & -20.2 & -25.5 & -26.2 & -18.4 \\ 
  $q_{50}$ & 17.4 & 9.4 & 6.6 & 5.6 & 3.6 & 2.2 & -0.7 & -3.5 & -4.0 & -0.3 \\ 
  $q_{84}$ & 148.7 & 82.0 & 56.9 & 43.5 & 35.5 & 27.4 & 22.3 & 22.7 & 17.7 & 14.4 \\ 
\hline
  \multicolumn{11}{c}{$\Delta Y/\Delta Z$ = 1}\\
  $q_{16}$ & -57.9 & -42.3 & -30.9 & -25.0 & -21.0 & -19.3 & -18.1 & -21.0 & -20.5 & -15.7 \\ 
  $q_{50}$ & 27.3 & 12.3 & 6.2 & 3.7 & 2.0 & 0.6 & 0.8 & 0.5 & -0.8 & -1.4 \\ 
  $q_{84}$ & 164.6 & 80.2 & 54.3 & 38.5 & 30.8 & 25.5 & 26.2 & 27.4 & 20.4 & 13.0 \\ 
\hline
  \multicolumn{11}{c}{$\Delta Y/\Delta Z$ = 3}\\
  $q_{16}$ & -70.7 & -69.1 & -44.0 & -32.1 & -26.0 & -20.4 & -19.4 & -20.6 & -16.6 & -9.9 \\ 
  $q_{50}$ & -19.2 & -7.5 & -3.6 & -0.5 & 0.0 & 0.5 & 0.8 & 2.0 & 2.1 & 2.0 \\ 
  $q_{84}$ & 85.3 & 49.9 & 35.9 & 31.0 & 26.9 & 23.7 & 24.4 & 29.3 & 22.9 & 14.8 \\ 
\hline
  \multicolumn{11}{c}{$\alpha_{\rm ml}$ = 1.50}\\
  $q_{16}$ & -17.4 & -13.9 & -9.1 & -7.8 & -7.2 & -6.6 & -6.4 & -8.5 & -8.7 & -9.0 \\ 
  $q_{50}$ & 88.1 & 47.8 & 32.3 & 24.2 & 18.1 & 15.5 & 16.1 & 16.1 & 8.8 & 2.0 \\ 
  $q_{84}$ & 289.7 & 154.8 & 105.2 & 79.0 & 66.7 & 57.9 & 58.5 & 51.4 & 37.4 & 17.9 \\ 
\hline
  \multicolumn{11}{c}{$\alpha_{\rm ml}$ = 1.98}\\
  $q_{16}$ & -78.2 & -84.8 & -76.7 & -58.3 & -47.7 & -40.0 & -34.1 & -33.5 & -30.2 & -22.3 \\ 
  $q_{50}$ & -54.0 & -40.2 & -26.2 & -18.0 & -14.2 & -12.1 & -11.0 & -10.7 & -6.5 & -4.4 \\ 
  $q_{84}$ & 38.6 & 19.0 & 13.6 & 11.7 & 9.9 & 8.9 & 9.2 & 12.6 & 11.3 & 7.3  \\ 
\hline
  \multicolumn{11}{c}{no diffusion}\\
  $q_{16}$ & -55.9 & -36.4 & -14.1 & -2.2 & 2.5 & 3.4 & 4.2 & 2.1 & -2.2 & -2.2 \\ 
  $q_{50}$ & 50.7 & 43.2 & 40.9 & 38.2 & 32.9 & 32.6 & 33.4 & 30.7 & 19.6 & 11.7 \\ 
  $q_{84}$ & 243.6 & 150.9 & 116.1 & 96.9 & 83.9 & 77.3 & 74.0 & 65.2 & 49.1 & 31.5 \\ 
\hline
  \multicolumn{11}{c}{standard, restricted to $M$ > 1.10 $M_{\sun}$}\\
  $q_{16}$ & -60.9 & -44.7 & -30.7 & -22.8 & -19.6 & -16.4 & -14.9 & -21.0 & -18.9 & -11.4 \\ 
  $q_{50}$ & 1.5 & 0.0 & 0.9 & 0.9 & -0.8 & -0.3 & 0.1 & 0.4 & 0.0 & -0.1 \\ 
  $q_{84}$ & 90.4 & 46.9 & 33.1 & 26.1 & 19.8 & 16.6 & 19.5 & 28.7 & 19.3 & 11.5 \\ 
\hline
  \multicolumn{11}{c}{overshooting $\beta$ = 0.2}\\
  $q_{16}$ & -60.0 & -43.7 & -32.3 & -24.6 & -22.1 & -21.6 & -20.1 & -16.9 & -1.1 & -21.9 \\ 
  $q_{50}$ & -1.4 & -3.5 & -4.5 & -4.2 & -6.4 & -8.5 & -6.2 & 5.7 & 16.5 & -6.0 \\ 
  $q_{84}$ & 78.9 & 38.8 & 23.3 & 16.9 & 10.8 & 8.4 & 21.9 & 30.7 & 34.4 & 14.6 \\ 
\hline
  \multicolumn{11}{c}{overshooting $\beta$ = 0.4}\\
  $q_{16}$ & -62.1 & -44.5 & -34.0 & -28.3 & -26.9 & -27.9 & -25.9 & -15.2 & -16.2 & -41.2 \\ 
  $q_{50}$ & -7.5 & -12.2 & -11.8 & -11.2 & -13.9 & -16.1 & -5.3 & -0.4 & -2.8 & -26.1 \\ 
  $q_{84}$ & 57.2 & 25.7 & 13.7 & 7.4 & 2.4 & 4.8 & 14.4 & 14.9 & 10.8 & -7.2 \\ 
\hline
\end{tabular}
\tablefoot{Typical Monte Carlo relative uncertainty on $q_{16}$ and $q_{84}$ is about 5\%, while the absolute uncertainty on $q_{50}$ is about 0.5\%.}
\end{table*}

\begin{figure*}
\centering
\includegraphics[height=17cm,angle=-90]{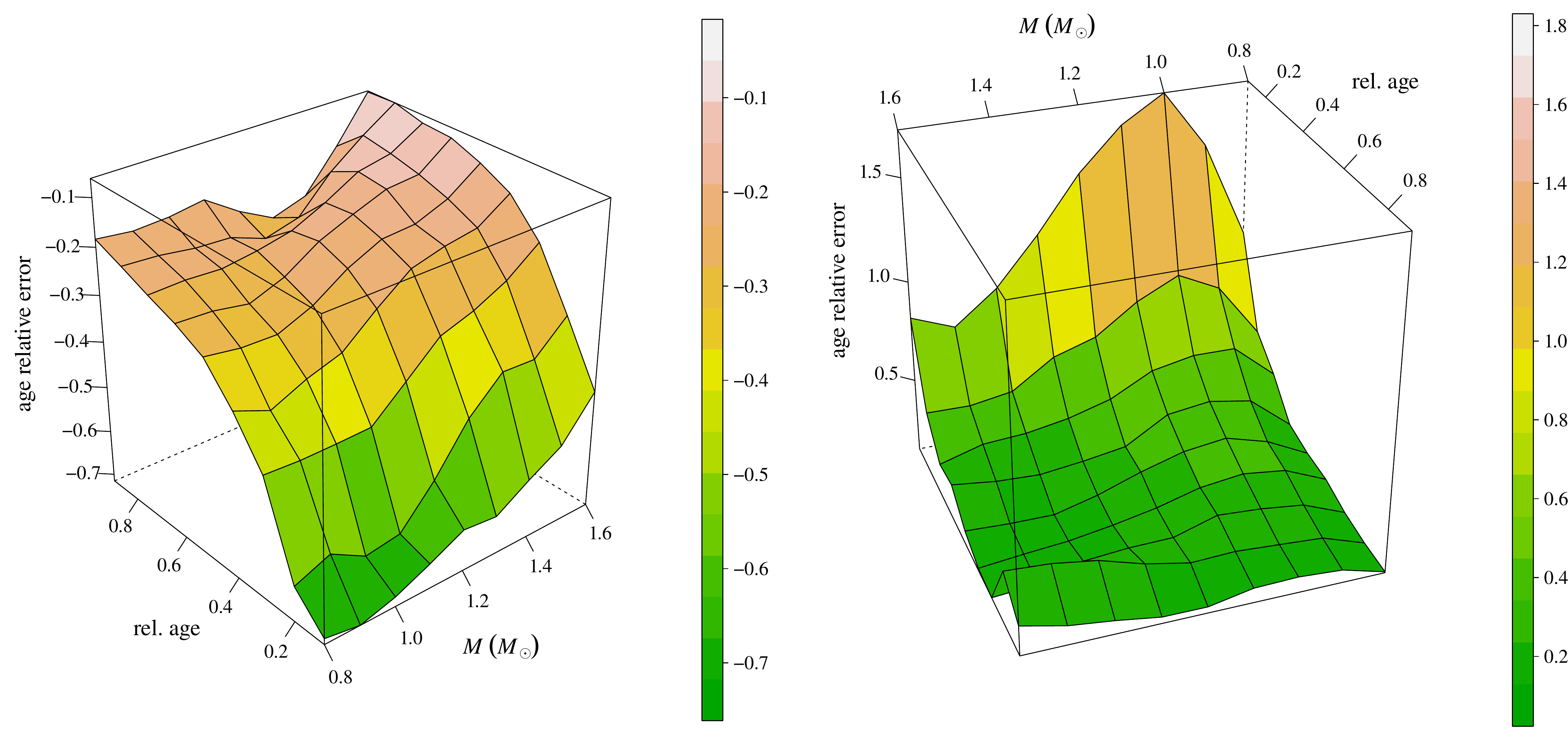}
\caption{{\it Left}: lower boundary of the 2D envelope of age estimate relative
  errors as a function of mass and relative age of the star. {\it Right}: same as
  in the {\it left panel}, but for the upper envelope boundary.}
\label{fig:relerror-2D}
\end{figure*}

Figure~\ref{fig:points} allows to assess the position of the $1 \sigma$
envelope boundary as a function of the mass of the star disregarding the
relative ages, or as a function of the stellar relative age disregarding 
the masses. 
To show how the mass and the relative age jointly influence
the boundary of the envelope, a 2D envelope was computed with a bidimensional
generalization of the technique described above\footnote{The width of the moving
windows in mass (0.07 $M_{\sun}$) and relative age (0.07) were chosen to
maintain a mean accuracy of about 1\% on the 2D Monte Carlo envelope.}.
The left panel of Figure~\ref{fig:relerror-2D} shows the position of the lower
boundary 2D envelope, while the upper one is in the right panel. Several
aspects discussed above are clearly visible; for example (right panel),
the edge effect at low mass and low relative ages
where we note the lack of models with overestimated age; this causes the strong decrease 
of the envelope in this region. 
The lower envelope boundary  (left panel) has a little decrease at about
1.3 $M_{\sun}$ at relative age greater than about 0.8. This effect is 
caused by the presence of a convective core for stars more massive
than 
about 1.1 $M_{\sun}$, which modifies the morphology of the stellar track and its
evolutionary time scale. The consequence is a displacement on the grid toward models 
of different mass and age. In particular
the relative age difference of the models around a 1.3 $M_{\sun}$ has a lower
16th quantile than 
that of less massive models. For target of about 1.5 $M_{\sun}$ an edge effect
masks the shift of the quantile since higher mass and lower age models are
under represented in the neighbourhood of the target point.

The results presented above can be compared with those of a similar analysis conducted by
\citet{Gai2011}. In that paper, which assumes the same uncertainties in the
observational constraints adopted here, the relative error in age
estimates are analysed. The results are presented as histogram and the
half-width at 
half-maximum (HWHM) is adopted to describe the spread of the distribution. An
overall HWHM of about 15\% is reported in \citet{Gai2011}. For comparison, we
computed a kernel density estimate \citep[see][and the Appendix A in
  \citealt{scepter1}]{Scott1992, 
  venables2002modern} for our results founding an HWHM
of about 20\%. A comparison of the two numbers should take into account that
the grid used in \citet{Gai2011} covers a different range of masses (up to
3.0 $M_{\sun}$) and includes red giants models. 
Moreover, Fig.~23 by \citet{Gai2011} shows a
variation in the HWHM according to the evolutionary phase of the stars: stars in
the earlier stages of evolution have a HWHM of their relative error histogram  
much larger than evolved stars. This is the same qualitative trend we analysed
in detail and quantitatively report
in Tab.~\ref{tab:global-pcage}.

\subsection{Impact of the observational errors and mass-age correlation}\label{sec:obs-err}

\begin{figure}
\centering
\includegraphics[height=8cm,angle=-90]{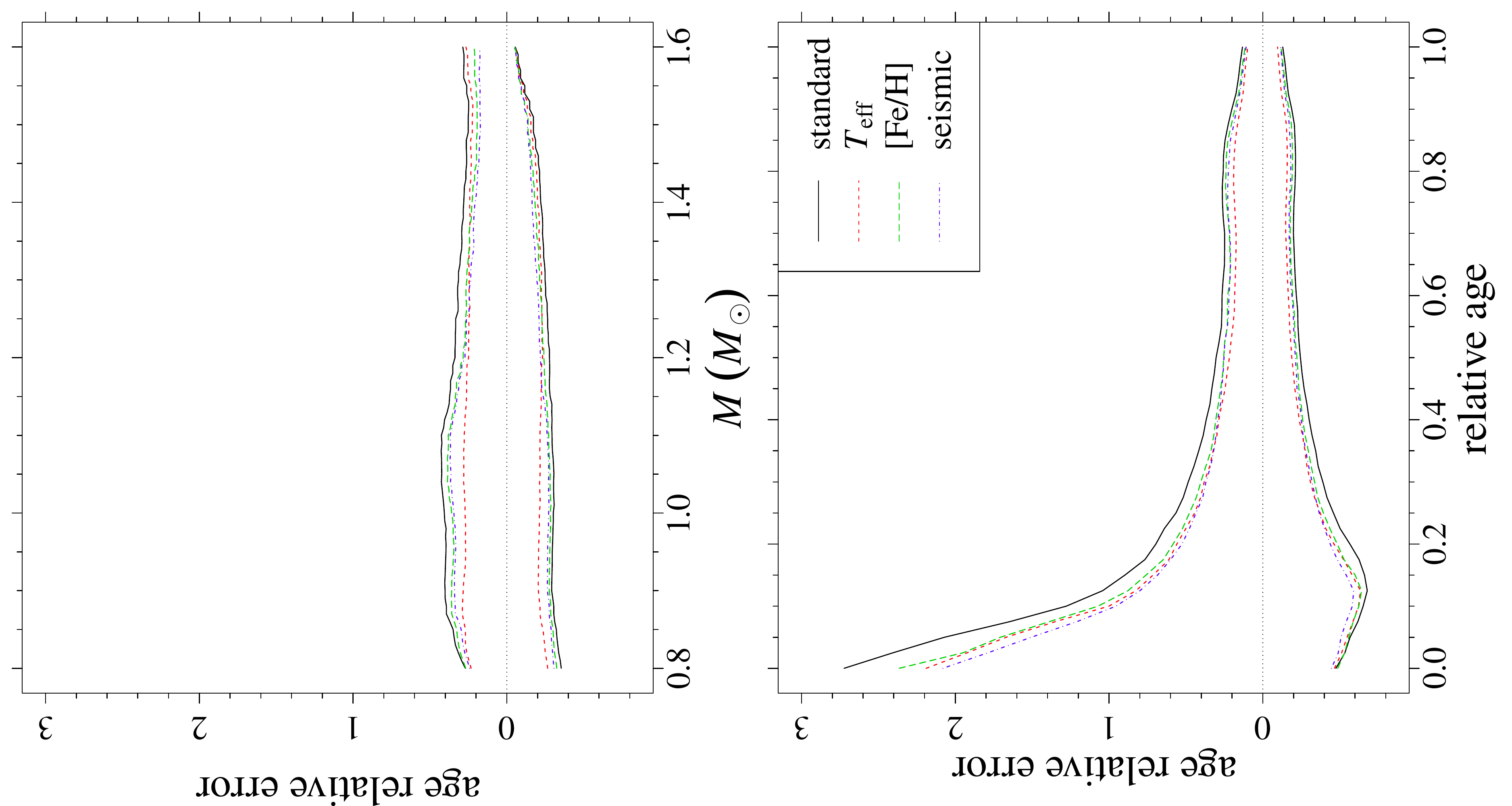}
\caption{Envelope of age estimate relative errors as a function of mass
  and relative age of the star for different assumptions on
  the observational errors. The solid line corresponds to the standard case;
  the dashed one to $\sigma(T_{\rm eff})$ = 50 K; the dotted one to $\sigma({\rm
  [Fe/H]})$ = 0.05 dex; the dot-dashed one to $\sigma(\Delta \nu)$ = 1\%,
  $\sigma(\nu_{\rm max})$ = 2.5\%.} 
\label{fig:relerror-sensitivity}
\end{figure}

To explore the sensitivity of the uncertainty on grid-technique age estimates 
	to the precision
of the data, we repeated the procedure by varying the observational uncertainties. 
 In this test we halved one
uncertainty at a time, while keeping the uncertainties in the other quantities
fixed to the standard value.  The first test assumes 50 K as $T_{\rm eff}$
error, the second test adopts an uncertainty of 0.05 dex in [Fe/H], while the
last one assumes uncertainty of 1\% in $\Delta \nu$ and 2.5\% in $\nu_{\rm
  max}$.  The results are presented in Fig.~\ref{fig:relerror-sensitivity} and
in Tab.~\ref{tab:global} and \ref{tab:global-pcage} in the sections labelled 
``$\sigma(T_{\rm eff})$ = 50 K'', ``$\sigma({\rm[Fe/H]})$ = 0.05 dex'', and
``$\sigma(\Delta \nu, \nu_{\rm max})$ = 1\%, 2.5\%''. 

Observing the shrink of the relative error on age estimates envelopes, it is apparent that the 
refinement of $T_{\rm eff}$ determination is the most important factor for
stars of mass lower than about 1.2 $M_{\sun}$, while for more massive objects
the metallicity and asteroseismic refinement have more importance.  The
reduction of the single-source observational uncertainties have a maximum
impact of about 10\% in the absolute shrink that is, about one-third
of the reference envelope half-width. Regarding the error envelope in
dependence on relative age, 
for relative ages lower than about 0.4 the seismic 
refinement causes the larger shrink of the envelope, while at
later evolutionary stages the effective temperature refinement is the most
important factor.

\begin{figure}
\centering
\includegraphics[height=8.5cm,angle=-90]{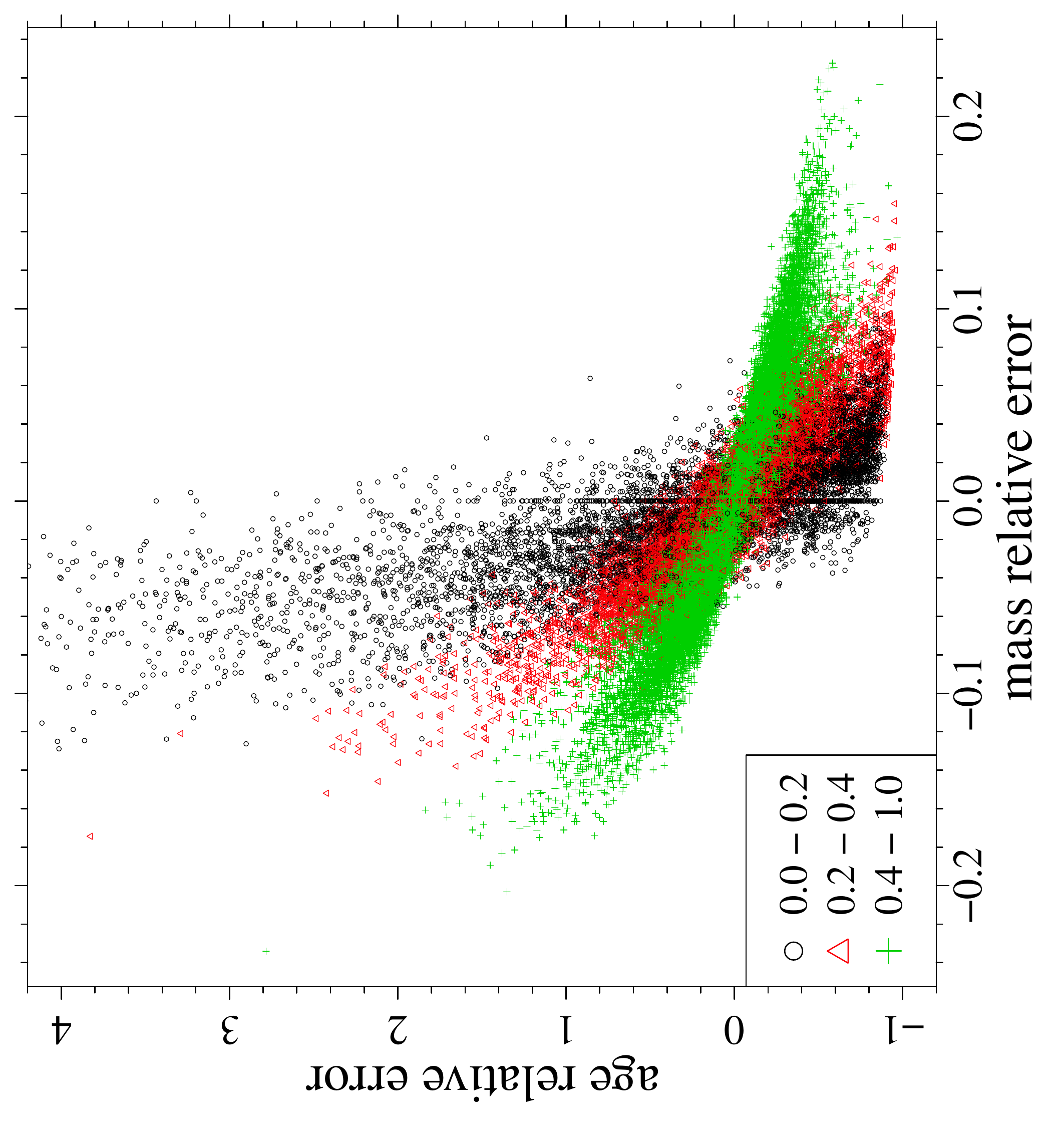}
\caption{Correlation between age and mass relative errors. The black circles
  correspond to models with relative age in 
  the range [0.0 - 0.2], the red triangles to relative ages in the range [0.2
    - 0.4], and the green crosses to relative ages in the range [0.4 - 1.0].}
\label{fig:corr_m_age}
\end{figure}

The errors on mass and age estimates are expected to show a negative
trend, because a star can be confused in the recovery procedure
either with an higher mass and lower age model or with a lower mass and higher
age 
one (see 
e.g. Fig.~4 in V14). It also follows from the discussion in V14 that the
strength of the linear correlation between mass and age relative errors
is expected to increase with relative age
of the 
star.  This trend is evidenced in Fig.~\ref{fig:corr_m_age}, which shows the
dependence between relative errors on mass and age estimates, grouped in
three relative age classes: models with relative age lower than 0.2; models
with 
relative ages between 0.2 and 0.4; models with relative age greater than 0.4.
The decrease of the slope of age versus mass relative errors with the
  increase of relative age is due to  
the mild increase of mass estimates variance at higher
relative age extensively discussed in V14, and the corresponding strong
decrease in variance in age estimates reported above. The correlation
coefficient (a measure of the dispersion of the data around the ideal linear
fit) between 
age and mass relative error in the three relative age groups are respectively
$-0.729$ 
(95\% confidence interval 
[$-0.740$ - $-0.718$]), $-0.906$ (95\% confidence interval [$-0.910$ -
  $-0.902$]), and $-0.922$ 
(95\% confidence interval [$-0.924$ - $-0.921$]). Estimated ages can never be
negative so the 
figure clearly shows the presence of an hard boundary at $-1.0$ for age
relative error.

\subsection{Impact of weighting the estimation grid}\label{sec:weight}

\begin{figure}
\centering
\includegraphics[height=8cm,angle=-90]{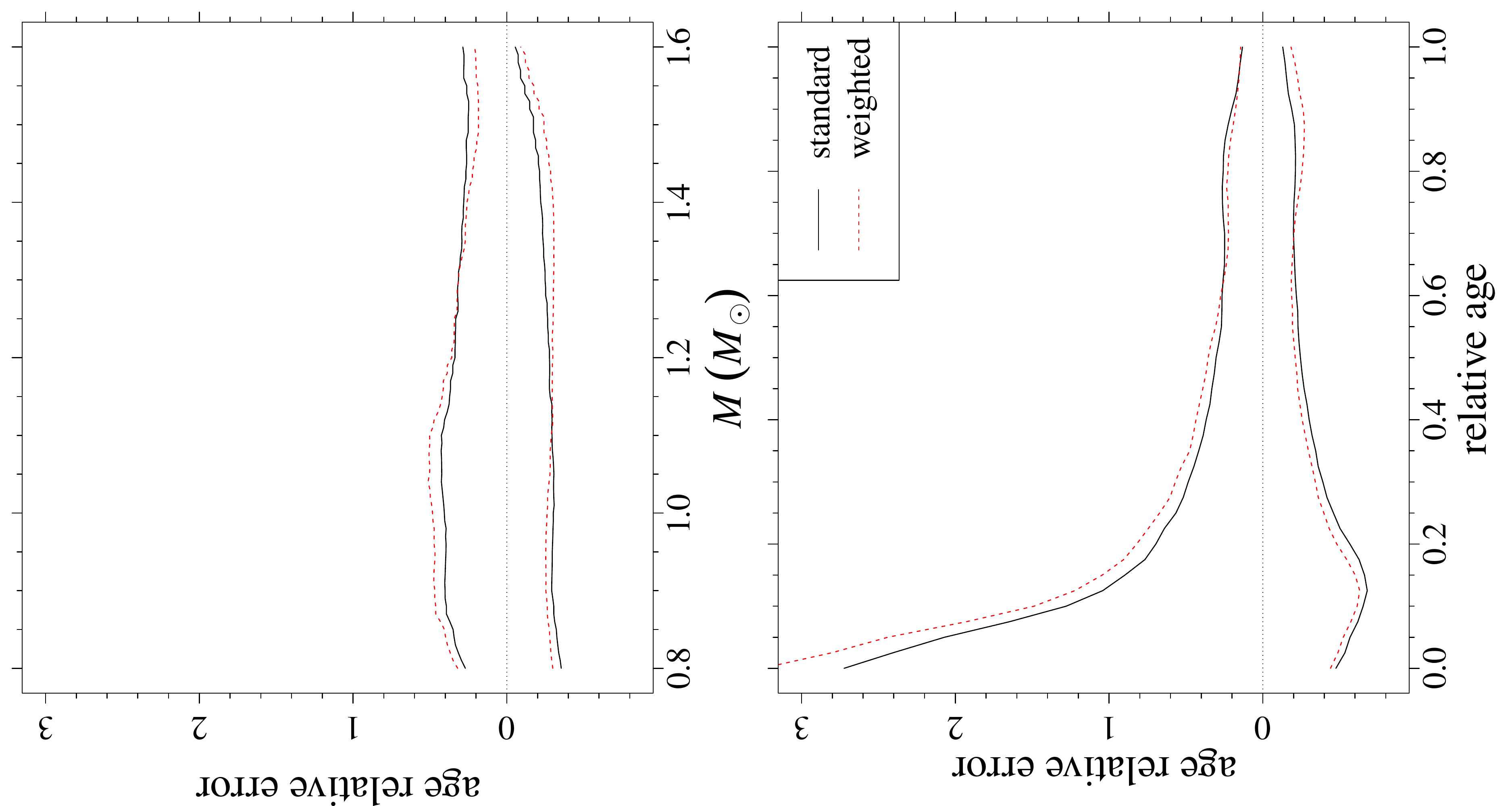}
\caption{Same as in Fig.~\ref{fig:relerror-sensitivity}, but for the comparison
of standard and weighted estimates (see text for details).}
\label{fig:relerror-weight}
\end{figure}

As a last check on the standard grid, we verified the influence of taking into
account the evolutionary time scale in the grid-based age estimation.  In fact
the grids of stellar models are biased toward rapid evolving stages, where
more points are computed to accurately follow the evolution.  It is well known
that the neglecting of this effect can lead to significant biased estimates
\citep[see e.g.][]{Jorgensen2005, Pont2004, Casagrande2011}.  
To quantify this effect,
we repeated the estimates described before on the same sample used in
  Sect.~\ref{sec:results} -- which was obtained
  sampling from the grid without 
taking the evolutionary time scale into account -- but recovering ages adopting
as a weight of each grid point the corresponding evolutionary time scale.  
The results are
summarised in Tab.~\ref{tab:global} and \ref{tab:global-pcage}, in the section
labelled "weighted'', and in Fig.~\ref{fig:relerror-weight}. It appears that
weighted estimates are slightly biased ($q_{50}$ about 3\%) toward ages higher
than those obtained adopting the standard grid for stars of mass lower
than about 1.1 $M_{\sun}$, while the 
opposite ($q_{50}$ about $-3\%$) occurs at higher masses. 
The same phenomenon
occurs in dependence on relative age with an age overestimation at lower
relative ages and an underestimation at high relative ages. These biases are
small with respect to the envelope width due to random errors.

The weights cause a slight underestimation of the mass for
  stars less massive than about 1.1 $M_{\sun}$, and a more pronounced
  overestimation for objects at about 1.4 $M_{\sun}$.
For low-mass
stars, whose tracks evolve nearly parallel in the ($T_{\rm eff}$, $\Delta
\nu$) plane, the weights have the largest influence in the first stages of
stellar 
evolution, until about relative age 0.6, where the differences in time scales
among models of different mass became smaller.

On the contrary, for more massive stars the overestimation in mass is
  due to the occurrence of tracks crossing during the overall contraction phase.
At the crossing, the time scale of the model of lower mass is shorter than that
of the more massive one, and therefore  the weighting estimates are biased
toward lower age.

\begin{figure}
\centering
\includegraphics[height=8cm,angle=-90]{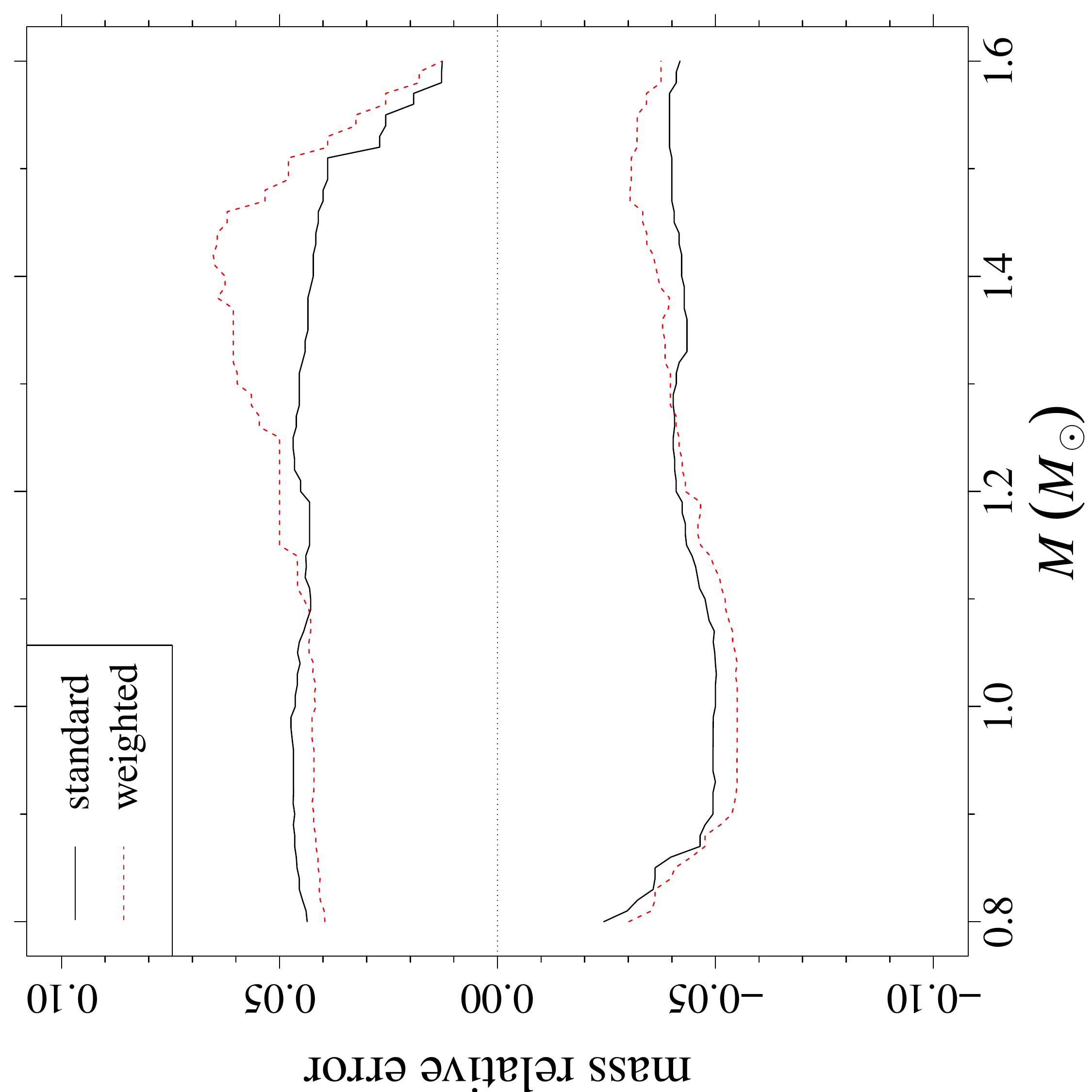}
\caption{Envelope of the mass relative error for stars sampled and
  reconstructed on the standard grid (black solid line), and for the same
  sample reconstructed on the weighted grid (red dashed line).}
\label{fig:relerror-M}
\end{figure}

Figure~\ref{fig:relerror-M} illustrates the effect showing the $1 \sigma$
  envelopes for mass 
  relative error with respect to the mass of the star for standard and
  weighted estimates. The bias occurring at high mass is apparent.

In this section we focused on the relative bias of the weighted age
  estimates 
  with respect to the standard one, considering weights only in the
  recovery stage. This quantified a maximum possible distortion of the
  unweighted estimates with respect to the weighted one.
  It seems that in presence of asteroseismic constraints weighted 
  estimates of mass presents a bias, especially for massive models, 
  not present in the unweigthed approach. 
Moreover, the differences of age estimates due to the grid weighting is much smaller than the random $1
\sigma$ envelope half-width. 
Therefore in
the following we adopt unweighted estimates as our reference scenario when
studying the effect of varied input in the stellar evolutionary computations.
We instead adopt weighted estimates in Sect.~\ref{sec:comparison} when
comparing with other 
pipelines which actually take into account such a correction.

\section{Stellar model uncertainty propagation}
\label{sec:errorprop}

When grid-based techniques are applied to real stars rather than 
to synthetic ones, the accuracy and precision of age estimates depend on the 
goodness of the adopted stellar models. A change within the uncertainties of the input adopted 
in stellar evolutionary codes directly propagates into a variation in the 
grid-based results. In V14, we discussed extensively this issue for mass and 
radius estimates. Here we perform a similar analysis
for age estimates. We focus our analysis on radiative opacity
uncertainty, on the value of the mixing-length parameter, on the initial
helium abundance, on the extension of any additional mixing region
starting from the border of the convective zones defined by the Schwarzchild
criterion, and on the efficiency of element diffusion. 

We performed these estimates following V14. More in detail, for each of 
the previously mentioned input we computed two non-standard grids of 
perturbed stellar models by varying the chosen individual input to its 
extreme values, while keeping all the others fixed to their reference values.
Artificial stars are then sampled from these grids and their ages are
estimated on the standard one. As in V14, we followed a slightly different procedure to study 
the effect of diffusion.  

The analysis of the difference between reconstructed and true values will quantitatively assess 
the effect of the quoted sources of uncertainties affecting modern stellar models. 

In general, perturbing a stellar model input leads to a twofold effect on the artificial stars. 
First, a displacement in the 4D space of the observational parameters 
(i.e. $T_{\rm eff}, {\rm [Fe/H]}, \Delta \nu, \nu_{\rm max}$) with 
respect to the location of standard models of the same mass and age. Second, a variation 
of the evolutionary time scale. As detailed in the following, there are cases in which the two effects are 
opposite and of the same order, thus resulting in a small net bias. In others, the latter effect is negligible and 
no compensation occurs, leading to a large bias. 

Notice that, given the strong dependence on the evolutionary phase of the age relative error, 
the bias at a given relative age must always be compared with the $1 \sigma$ envelope at the same phase 
to assess the relevance of the bias itself.

\subsection{Initial helium abundance}
\label{sec:he}

The helium-to-metal enrichment ratio $\Delta Y/\Delta Z$, commonly adopted 
by stellar modellers to select the initial helium abundance, is quite uncertain 
\citep{pagel98,jimenez03,gennaro10}. To quantify the impact of such an uncertainty on 
grid-based age estimates, we computed two additional grids of stellar models with
the same values of the metallicity $Z$ as in the standard grid, but by
changing the helium-to-metal enrichment ratio $\Delta Y/\Delta Z$ to values 1
and 3. Then, we built two synthetic datasets, each of $N = 50\,000$ artificial stars,
by sampling the objects from these two non-standard grids. The age of the
objects are then estimated using the standard grid for the recovery.  

\begin{figure*}
\centering
\includegraphics[height=6cm,angle=-90]{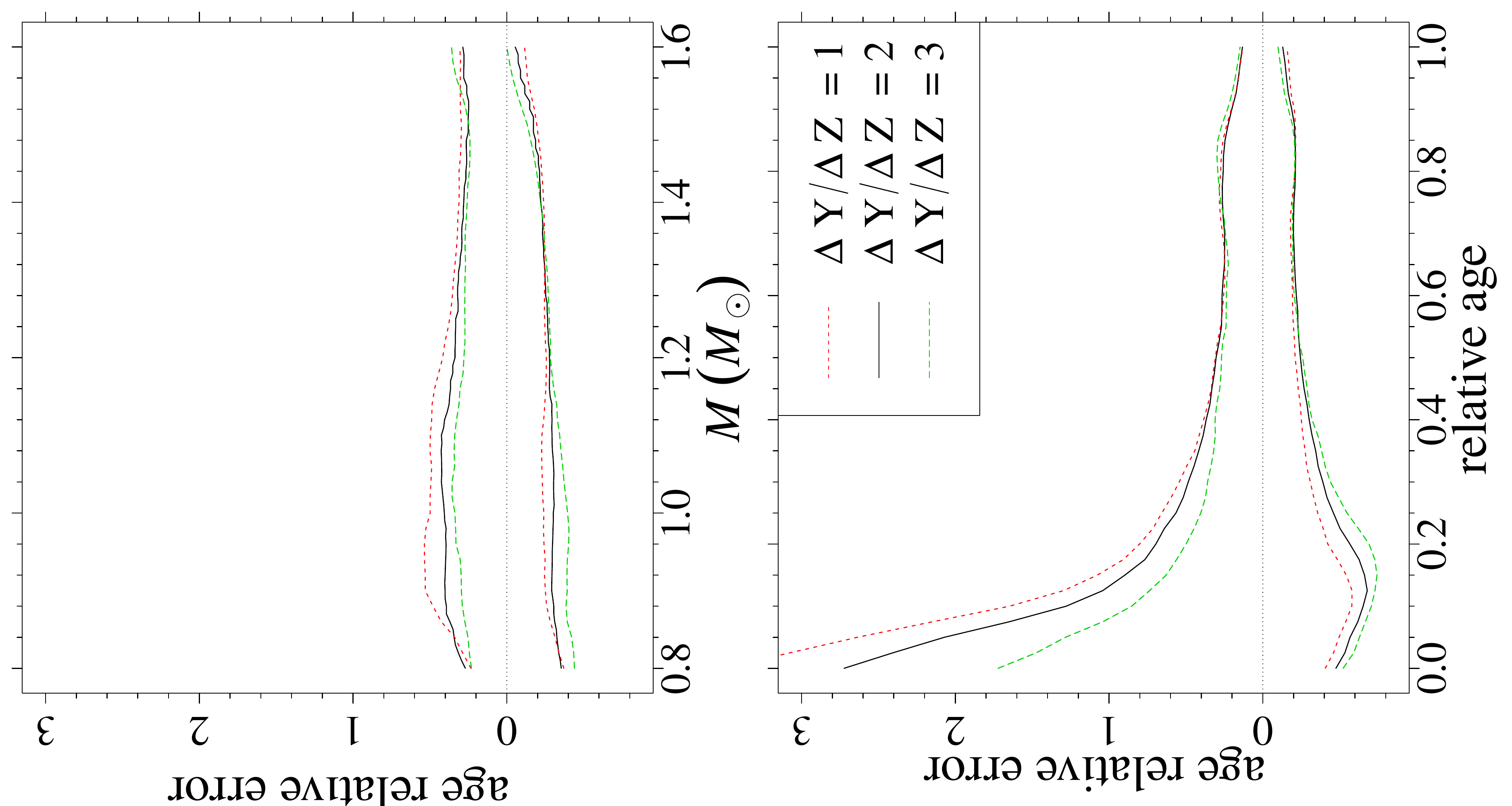}
\includegraphics[height=6cm,angle=-90]{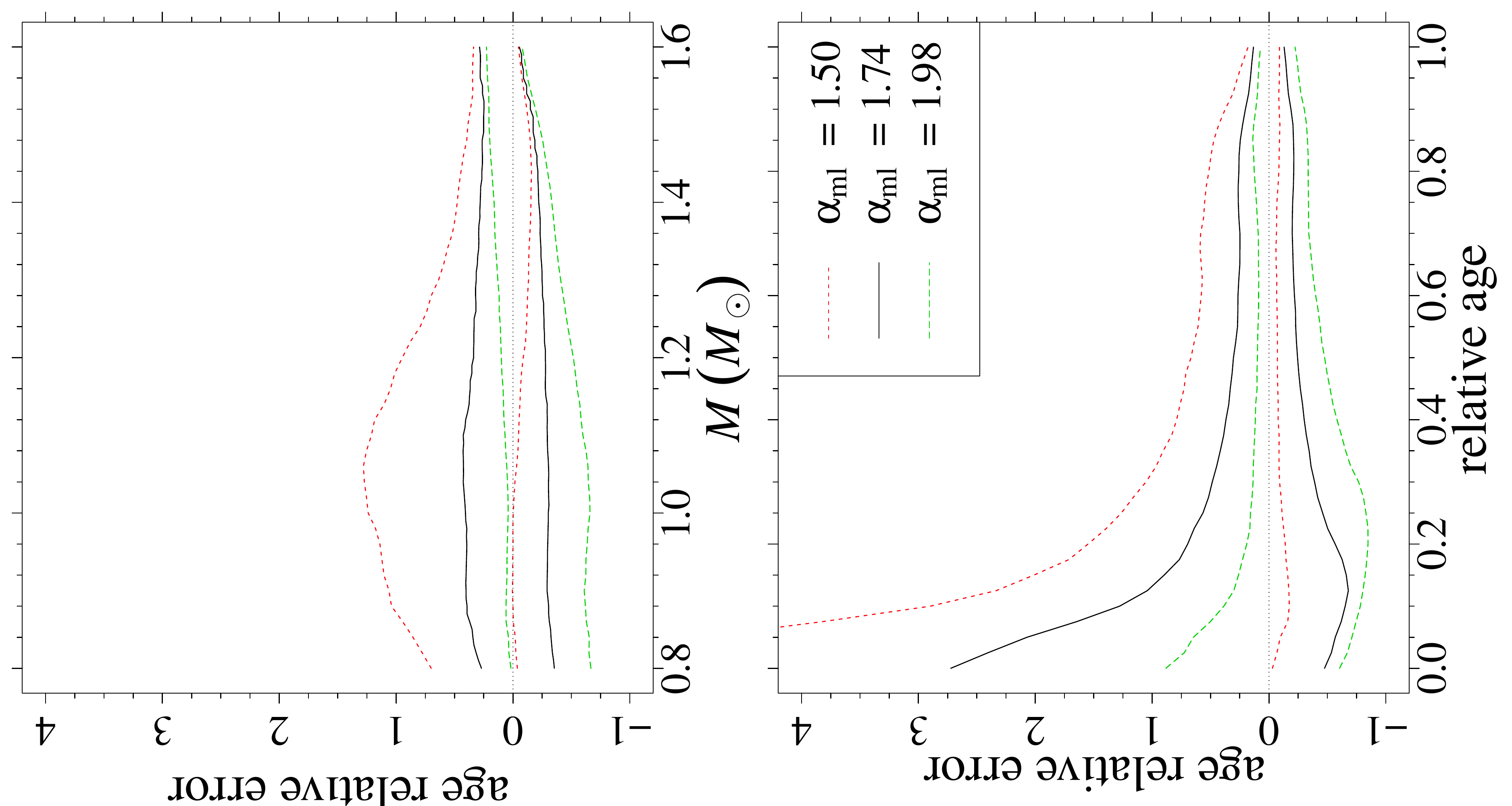}
\includegraphics[height=6cm,angle=-90]{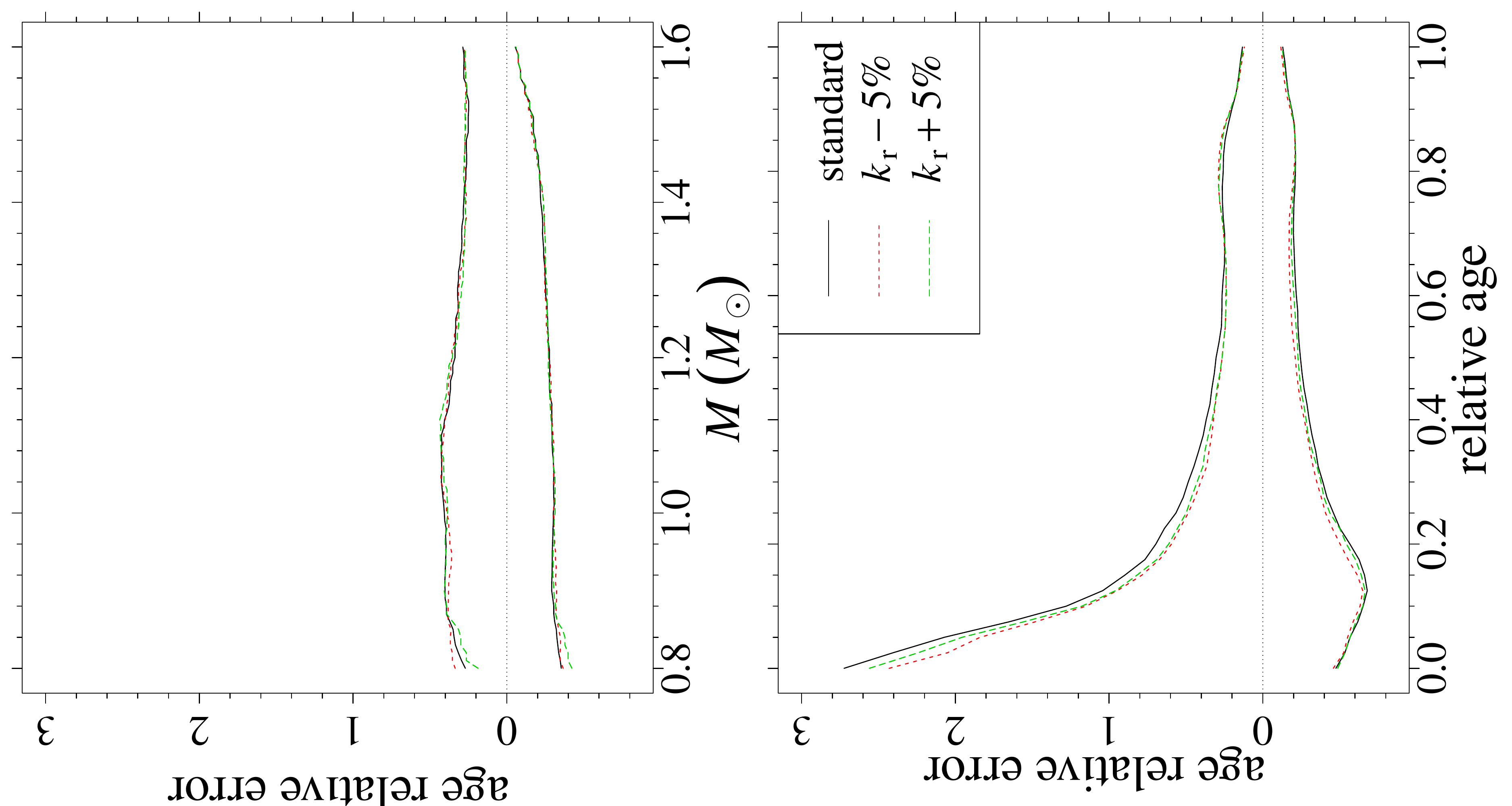}
\caption{{\it Left column}: envelopes of age estimate relative errors due to sampling from synthetic
  grids with different values of $\Delta Y/\Delta Z$. Ages are estimated on the
  standard grid with $\Delta Y/\Delta Z$ = 2. {\it Middle column}: same as the
  {\it left column}, but for sampling from synthetic grids with different values of
  the mixing-length parameter $\alpha_{\rm 
    ml}$. Ages are  
  estimated on the standard grid with $\alpha_{\rm ml}$ = 1.74. {\it Right column}:
  same as the {\it left column}, but for sampling from synthetic grids with different values of radiative
opacity $k_{\rm r}$. Ages are estimated on the standard grid.}
\label{fig:relerror-He_ML_kr}
\end{figure*}

The results of these tests are presented in section "$\Delta Y/\Delta Z =
1$'' and "$\Delta Y/\Delta Z = 3$'' of Tables~\ref{tab:global} and
\ref{tab:global-pcage} and in the left column of
Fig.~\ref{fig:relerror-He_ML_kr}. 
The effect of the initial helium content change is on average modest. For $\Delta
Y/\Delta Z = 1$ a bias with respect to standard case median of about 3\% to 5\% occurs for stellar masses below
1.1 $M_{\sun}$, with a similar effect on the envelope boundaries. For $\Delta
Y/\Delta Z = 3$ the bias ranges from about -2\% to -5\% for stellar masses
below 1.1 
$M_{\sun}$, with an envelope shift from about -7\% to -10\%. 
The largest bias with respect to standard scenario results occurs at low relative 
age  where it reaches values
 of about 21\% and $-25\%$ for $\Delta
Y/\Delta Z = 1$ and 3.

\begin{figure}
\centering
\includegraphics[height=8cm,angle=-90]{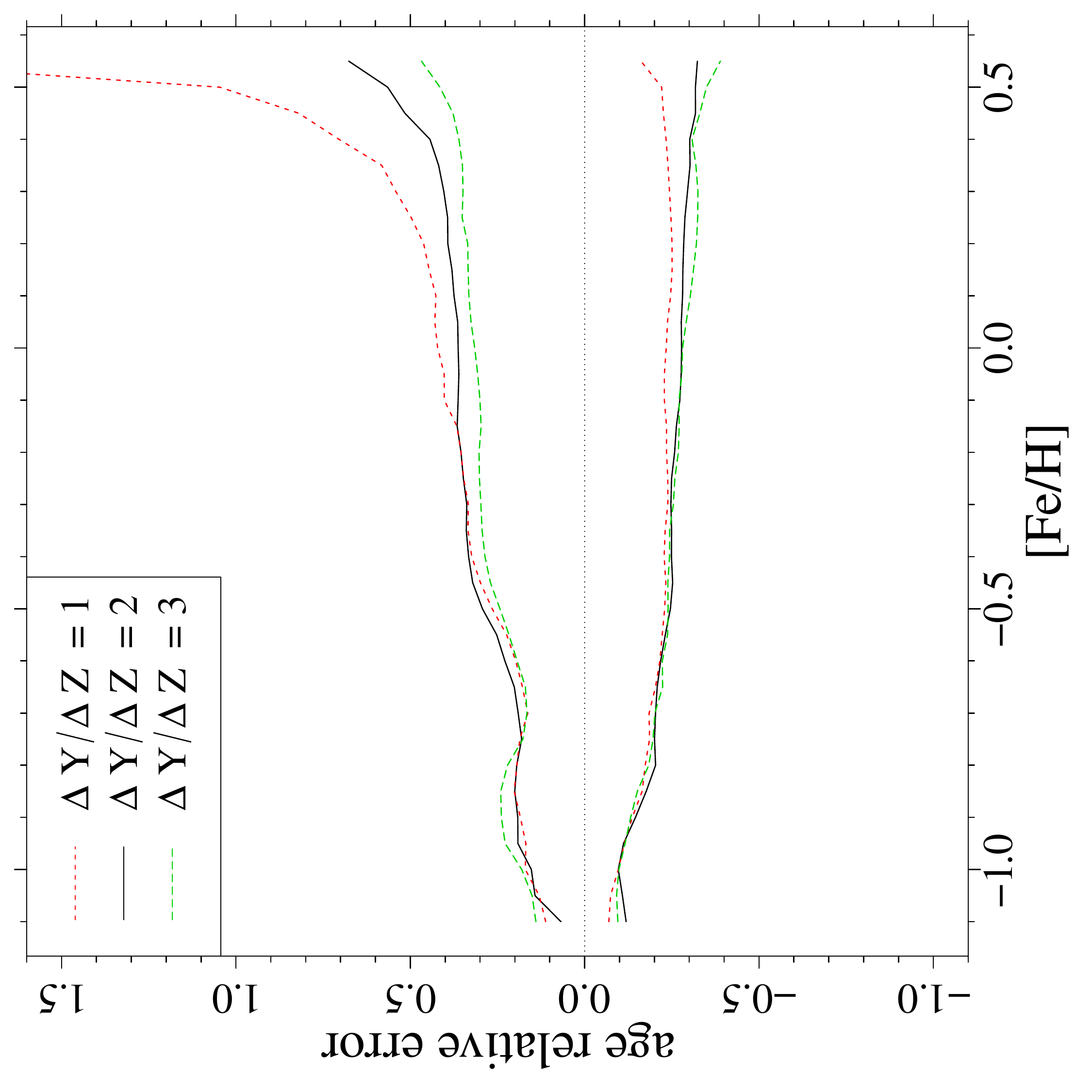}
\caption{Dependence on the surface metallicity [Fe/H] of the envelopes of age
  estimate relative errors. The sampling occurs from 
  synthetic 
  grids with different values of $\Delta Y/\Delta Z$, while ages are estimated
  on the 
  standard grid with $\Delta Y/\Delta Z$ = 2.}
\label{fig:relerror-feh_he}
\end{figure}

As shown in V14, the largest impact of the initial helium change
occurs, as expected, for high values of [Fe/H]. To illustrate this point, let
us consider the  
sampling from the grid with $\Delta Y/\Delta Z = 1$ and the reconstruction on
the 
standard grid ($\Delta Y/\Delta Z = 2$).
For stars at the upper metallicity edge the effect of helium change is
 relevant. Figure~\ref{fig:relerror-feh_he} shows the age relative error
envelope for the considered values of helium-to-metal enrichment ratio 
as a function of the surface [Fe/H] value. It is
apparent that the low helium scenario presents a very long tail towards age
overestimation near [Fe/H] = 0.50 dex. This is due to the fact that, for
 [Fe/H] values near to the edge boundary, the $\Delta Y/\Delta Z = 1$ models
 are significantly shifted toward higher asteroseismic parameters. 
 As a consequence they are often confused in the recovery 
 with standard near ZAMS models whose age are more difficult to estimate.

The overall smallness of the
effect of an helium abundance change might be
unexpected since  the 
initial helium content significantly affects the evolutionary time scale of
stellar models. 
Moreover in V14 we showed that the same variation in $\Delta Y/\Delta Z$
results in a sizeable uncertainty in grid-based recovery of $M$ and $R$.
This effect is also confirmed for the masses up to 1.6 $M_{\sun}$ studied
here.
The obtained results are due to the concurrent change of
the effective temperature and seismic parameters. This effect is 
evidenced in Fig.~\ref{fig:diff-grid}, which reports the boxplots of the
differences in effective temperature and age for stellar models of same
relative age   
computed with standard and varied input\footnote{
A boxplot is a convenient way to summarize the
variability of the data;  
the black thick lines show the median of the data set, while the box 
marks the interquartile range, i.e. it extends form the 25th to the 
75th percentile of data. The  whiskers extend from the box until the
extreme data,  but they can extend only
to a maximum of 1.5 times the width of the box.}. 
To enhance the figure
readability the outliers, which can be presented as individual points, are
omitted from the plot. 
The medians shift are related to the effect of the input
variation in the stellar evolutionary code. The larger the separation of the
medians  
with respect to the interquartile distance of the boxes and the greater
the  importance  of a given input variation.

 It is apparent that a change
in the initial helium content has a notable effect in both ages and effective
temperature variations and that such variations partially counterbalance
each other.  As discussed in V14, an artificial star with enriched initial helium will
  have -- at fixed evolutionary phase -- an higher effective temperature
  and a lower age than the corresponding model of the same mass but computed with standard initial
  helium abundance. The change in the observational parameters forces the
  helium-rich 
  star to lie in a zone of the standard grid populated by more massive models, leading to a mass overestimate in the recovery. 
However, helium-rich stars evolve faster than corresponding standard scenario stars.
It happens that the age bias due to the mass overestimate nearly compensates the difference in age due to the change in the initial 
helium, thus resulting in a small net
bias in estimated age.
It is also apparent that the balancing effect is more accurate for
  massive models and for later evolutionary stages, when the error envelopes
  computed with modified initial helium abundance converge to the standard one.

\subsection{Mixing-length value} 
\label{sec:ml}

It is increasingly apparent that the use of a solar-calibrated mixing-length
value for stars which differ from the Sun in mass, composition, and/or 
evolutionary phase could be not appropriate \citep[see e.g.][]{Deheuvels2011, Bonaca2012,
  Mathur2012,  Tanner2014, Trampedach2011, Magic2014, Yildiz2007, Clausen2009}.
 
To quantify the effect of varying the efficiency of the super-adiabatic convective transport,
 we computed two additional grids of stellar models by assuming
mixing-length parameters $\alpha_{\rm ml}$ = 1.50 and 1.98.  Then, we built two synthetic datasets, each of $N =
50\,000$ artificial stars, by sampling the objects from these two
non-standard grids. The age of the objects are estimated using the standard
grid for the recovery, which assumes  our solar-calibrated
value $\alpha_{\rm ml}$ = 1.74.  The results of these tests are presented in section
``$\alpha_{\rm ml} = 1.50$'' and ``$\alpha_{\rm ml} = 1.98$'' of
Tables~\ref{tab:global} and \ref{tab:global-pcage} and in the central column of
Fig.~\ref{fig:relerror-He_ML_kr}.

\begin{figure*}
\centering
\includegraphics[height=17cm,angle=-90]{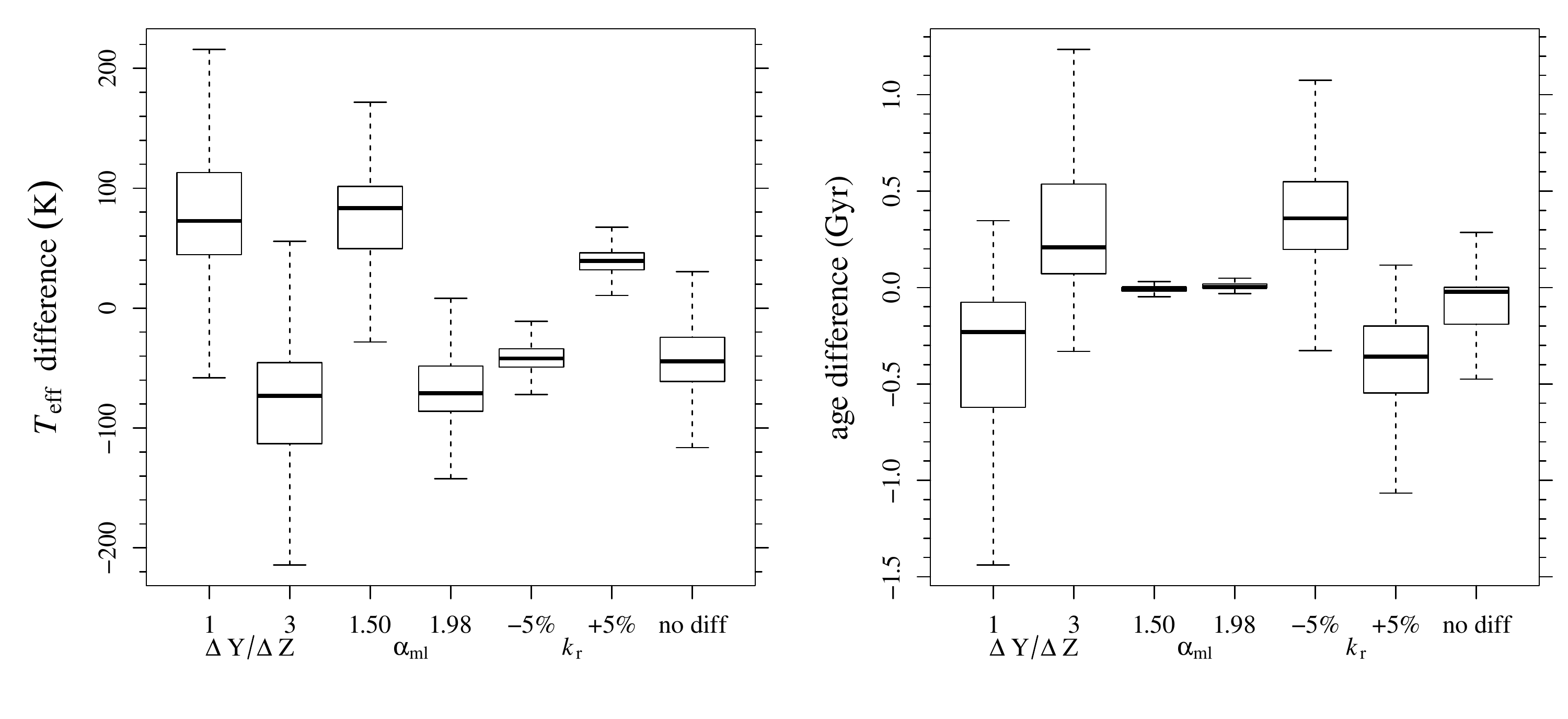}
\caption{{\it Left}: Boxplots of the difference in effective temperature between
  standard models and models with varied input or
  parameters. The differences are computed for models in the same evolutionary
  phase. A positive value implies that the standard grid provides an higher
  effective temperature. {\it Right}: same as  
  the {\it left panel}, but for age.
}
\label{fig:diff-grid}
\end{figure*}

In this case the bias is very large with values ranging from 20\% to 30\% for models
of mass lower than 1.2 $M_{\sun}$. As expected, the bias is lower for higher
mass models due to the decreasing thickness of the convective envelope.
The bias due to the mixing-length variation is nearly the same as the
1$\sigma$ random uncertainty of the standard models, implying that the age
estimates are prone to systematic biases due to the adoption of an improper
mixing-length value. The origin of this large bias can be understood with the
same argument discussed for initial helium content. 
Artificial stars with varied $\alpha_{\rm ml}$ occupy a different location 
in the 4D space of the observable quantities with respect to standard models 
of the same mass and age. Consequently the recovered mass and age on the standard grid will be necessarily biased. 
Moreover, contrarily to the case of a variation in the helium abundance, changing
 the mixing-length does not affect the evolutionary time scale (see the boxplots in 
Fig.~\ref{fig:diff-grid}). As a result, the counterbalancing effect previously described in Sect.~\ref{sec:he} 
can not occur and the age bias is large.

\subsection{Radiative opacity} 
\label{sec:kr}

In \citet{incertezze1,incertezze2}, we devoted a strong computational effort
to quantify the cumulative uncertainty affecting stellar models due to the
combined effects of the main input physics.  As a result we found that the main
source of variation is due to the current uncertainty in the radiative
opacities.

To quantify the impact of this uncertainty source in age estimates, we
computed two additional grids with values of radiative opacity increased and
decreased by 5\% (see the discussion in \citealt{incertezze1} for the choice
of the quoted uncertainty).  Then, we built two synthetic datasets, each of $N =
50\,000$ artificial stars, by sampling the objects from these two
non-standard grids. The age of the objects are then estimated using the standard
grid for the recovery.

The results of these tests are presented in the right column of
Fig.~\ref{fig:relerror-He_ML_kr}\footnote{Due to the negligible effect, we do not include the results in Tables~\ref{tab:global} and \ref{tab:global-pcage}.}.
Although a $\pm 5\%$ variation in radiative opacity has a large
influence on the evolutionary time scale of stellar models
\citep[][]{incertezze1,incertezze2}, it appears that such a change does not have
effect in grid-based age estimates. The reason for that is
the same kind of counterbalancing effect discussed in Sec.~\ref{sec:he} for the variation of initial helium content.

\subsection{Convective core overshooting} 
\label{sec:ov}

The lack of a self-consistent treatment of convection in stellar 
model computations prevents a firm prediction of the convective 
core extension. The usual approach consists in parametrizing the extension 
of  the extra-mixing region beyond the canonical border, as defined by the 
Schwarzschild criterion, in terms of the pressure scale height $H_{\rm 
  p}$: $l_{\rm ov} = \beta H_{\rm p}$, where $ \beta $ is a free parameter. 
  To quantify the impact of taking convective core overshooting into account we
computed -- only for models more massive than 1.1 $M_{\sun}$ --  two additional
grids with values of $\beta$ = 0.2 and 0.4, the last representing a possible
maximum value for the overshooting extension \citep[see e.g. the discussion
  in][]{cefeidi}.  
Then, we built two synthetic datasets, each of $N =
50\,000$ artificial stars, by sampling the objects from these two
non-standard grids. The age of the objects are then estimated using the standard
grid -- restricted to models more massive than 1.1 $M_{\sun}$ -- for the
recovery. 

\begin{figure}
\centering
\includegraphics[height=8cm,angle=-90]{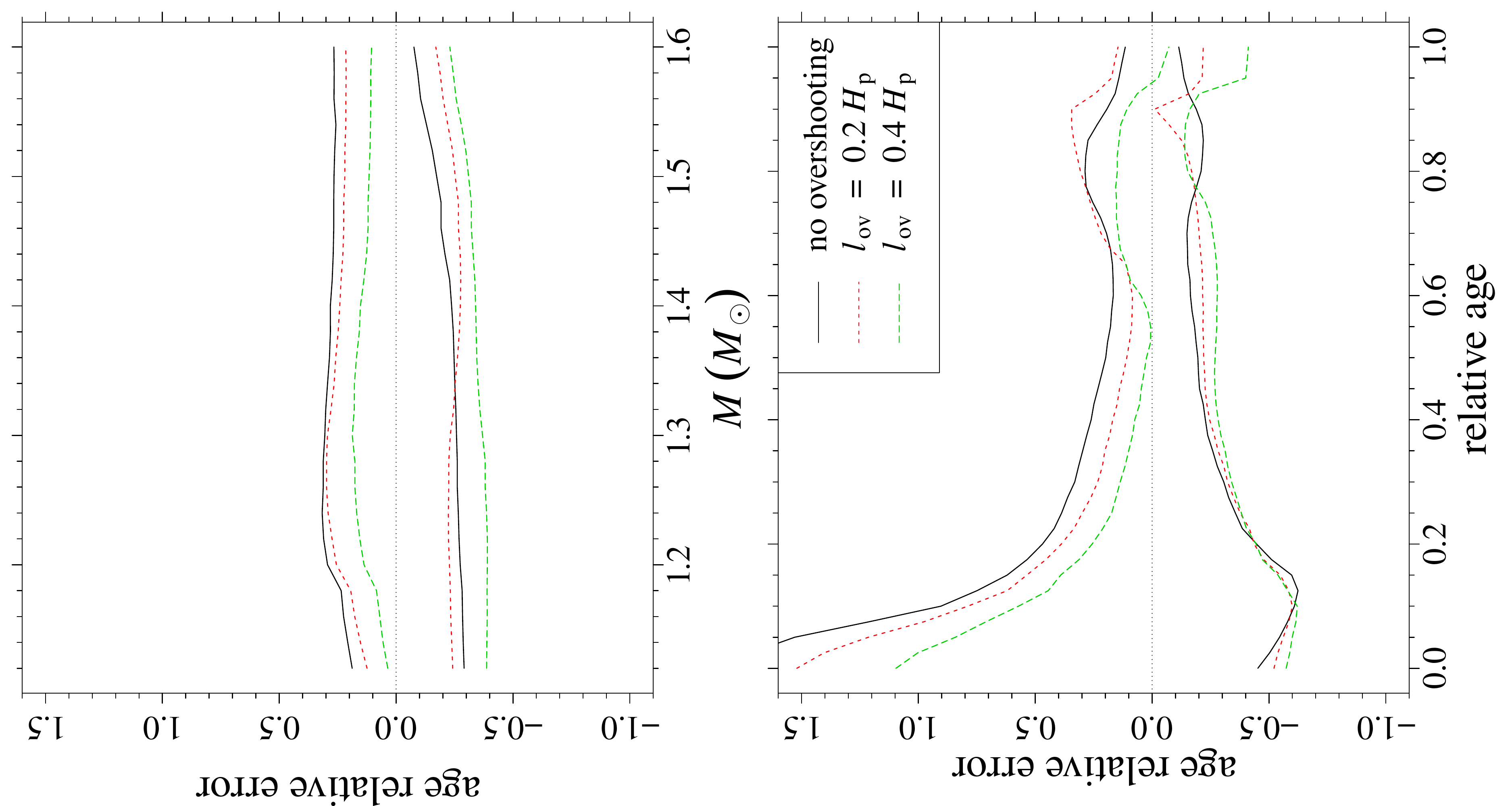}
\caption{Envelope of age estimate relative errors due to sampling from synthetic
  grids with different values of convective core overshooting extension $l_{\rm ov}$. Ages are estimated on the
  standard grid without overshooting.}
\label{fig:relerror-OV}
\end{figure}

The results of these tests are presented in section
"overshooting $\beta = 0.2$'' and "overshooting $\beta = 0.4$'' of
Tables~\ref{tab:global} and \ref{tab:global-pcage} and in 
Fig.~\ref{fig:relerror-OV}.
The bias due to the mild-overshooting scenario is at most about $-7\%$ for models
of 1.5 $M_{\sun}$, while for $\beta = 0.4$ it reaches values of about $-13\%$.

\begin{figure}
\centering
\includegraphics[height=8cm,angle=-90]{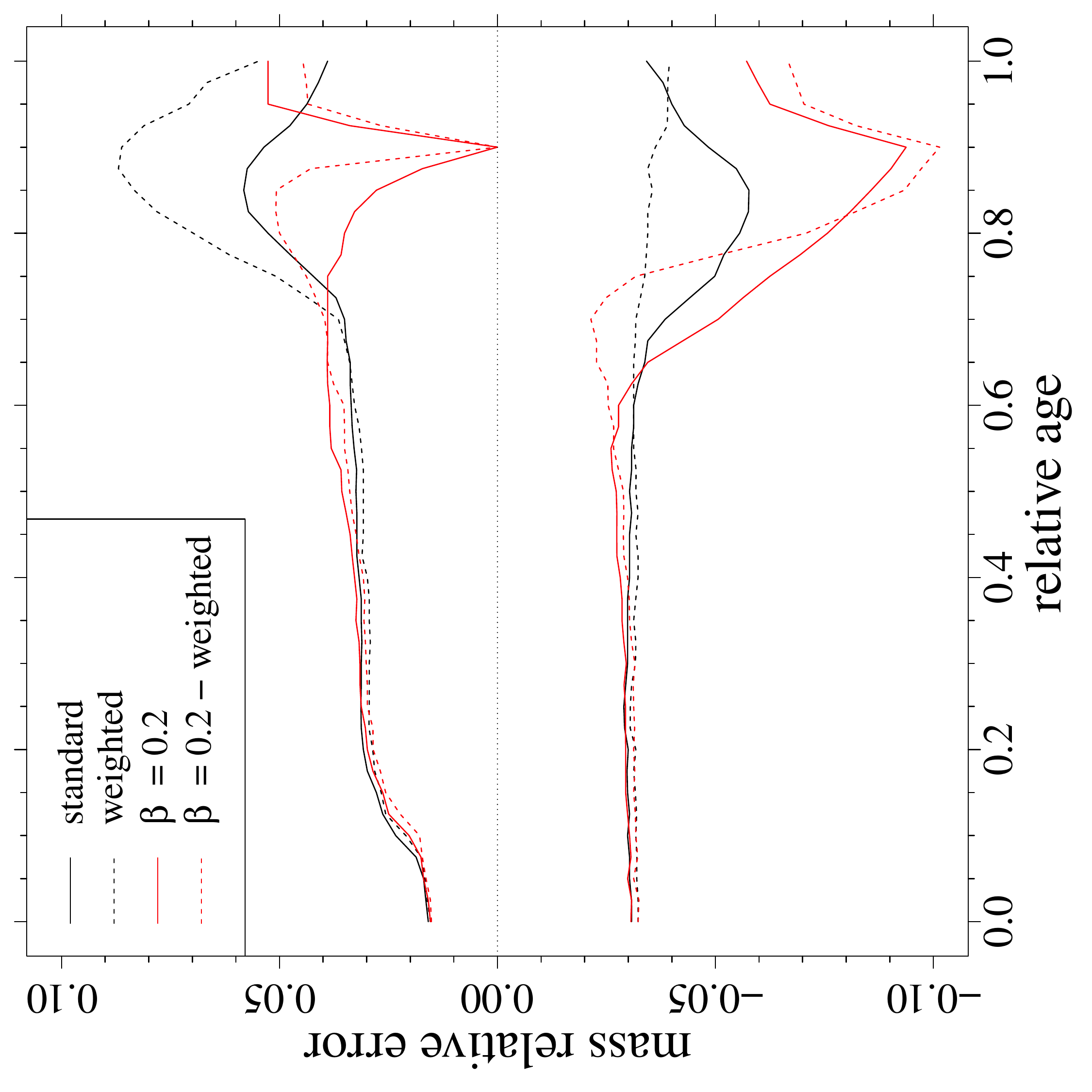}
\caption{Envelope of mass estimate relative errors for the reference case
  (black solid line), and for the weighted estimates (dashed black line).
The red lines correspond to the overshooting scenario with $\beta = 0.2$ (red
solid line) and
of the same scenario adopting weighted estimates (red dashed line). 
}
\label{fig:relerror-OV-M}
\end{figure}

As usually happens, the bias in age estimate is essentially 
	the consequence of that in mass.	 
	In V14 the effect of overshooting on mass estimates was not
	studied since the range of mass was different, so we present here the basic
	results.
	 Figure~\ref{fig:relerror-OV-M}
	shows the effect on mass estimates of a mild overshooting $\beta = 0.2$. 
	The effect of the overshooting is evident around relative ages from 0.7 to 0.9.
In this zone the morphology of the grids computed with and without overshooting is most different,
since the overall contraction starts in different regions of the two grids.
This cause the strong bias toward mass underestimation visible in  Fig.~\ref{fig:relerror-OV-M}. When the overall contraction phase ends
and the tracks again evolve parallel (relative age around 0.9), the bias suddenly disappears.
 
Since the largest bias occurs in
the same region where the impact of  weighted estimates is higher, we also show in
the figure the weighted estimate envelopes for standard and mild
overshooting 
scenarios. As discussed above, the weighting biases the mass estimates toward
higher values; the effect partially counterbalances -- in the relative age range
from 
0.7 to 0.8 -- the effect of the overshooting, resulting in a final less biased
estimates.

Turning again to age, the trends in the relative age error envelopes in Fig.~\ref{fig:relerror-OV} show 
	the signature of the mass bias discussed above.
We also note that around relative age 0.8, the standard envelope of the relative age errors shows an inflation, caused by the
degeneracy present in the grids during the overall contraction phase (see
Sect.~\ref{sec:results}).  
Since several
tracks accumulate and cross in this zone, the grid estimation procedure is here
intrinsically more difficult. 
At relative age of about 0.9 the
age estimates for the overshooting scenarios show a sudden bias toward lower ages. 
Such an occurrence can be understood by remembering that models with
convective core overshooting evolve slower than standard models until this
phase, and faster after.

\subsection{Elements diffusion} 
\label{sec:feh}

In V14 we discussed the importance of taking into account the effects of the
microscopic diffusion when determining stellar parameters by means of
grid-based techniques, assessing the bias in mass and
radius estimates when element diffusion is neglected.

A similar analysis was performed here for age estimates. As in V14, we followed
 an approach slightly different from that of the
previous Sections. We think that it is more realistic in this case to build the
artificial stars by sampling $N = 50\,000$ objects from the standard grid of
models, which takes the element diffusion into account, and to use
non-standard models, which neglects elements diffusion, for the recovery.

\begin{figure}
\centering
\includegraphics[height=8cm,angle=-90]{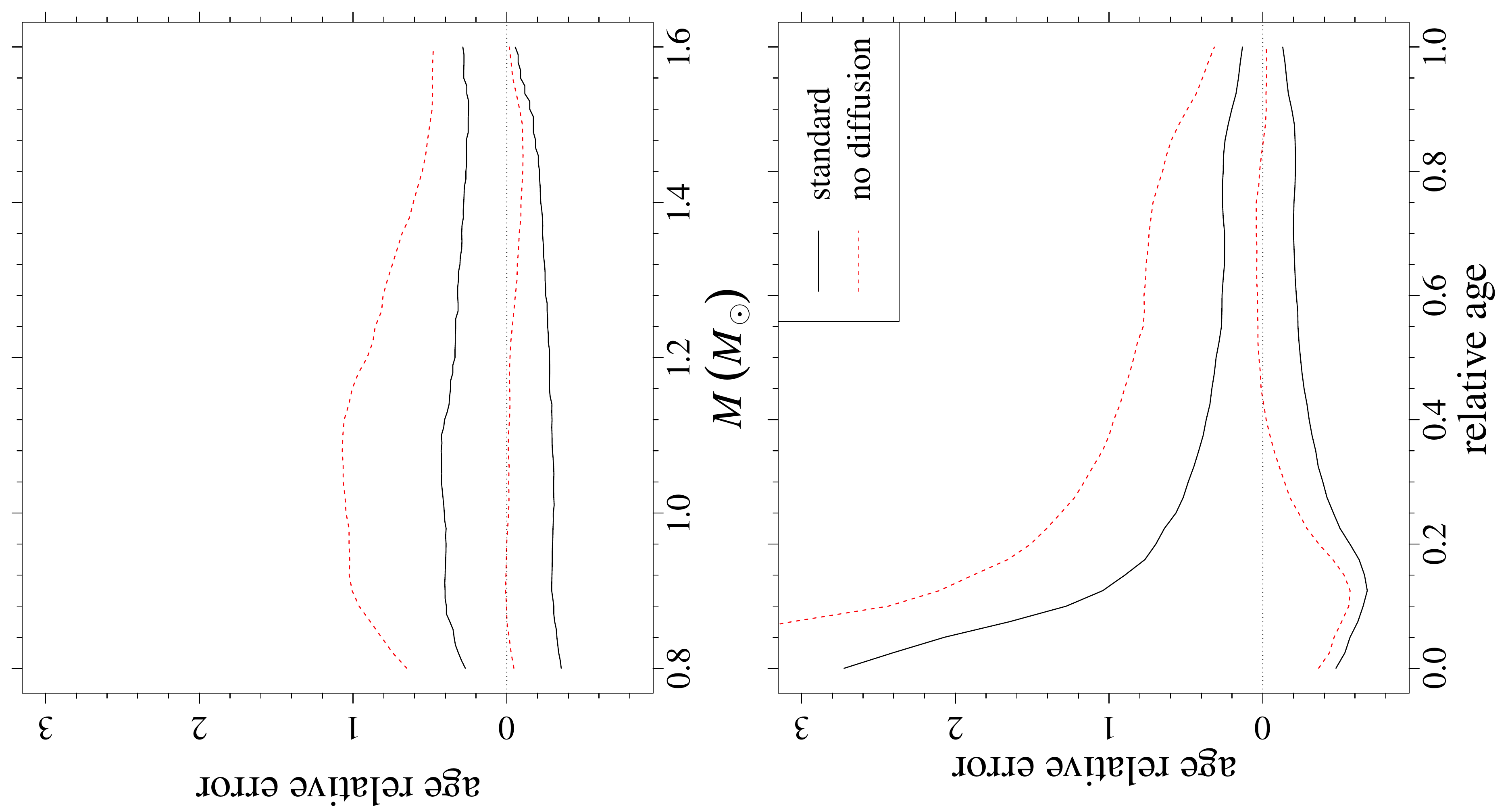}
\caption{Envelope of age estimate relative errors obtained by 
including (standard) or neglecting diffusion in the reconstruction grid.
At variance with previous figures, in this case the synthetic dataset is
sampled from the standard grid.
}
\label{fig:relerror-diff}
\end{figure}

The results of the test are presented in section ``no diffusion'' of
Tables~\ref{tab:global} and \ref{tab:global-pcage} and in
Fig.~\ref{fig:relerror-diff}.

In this case the bias is very large, reaching values of about 40\% for stars
of mass lower than 1.1 $M_{\sun}$. As expected the bias is lower for more massive 
objects due to the fast evolutionary time scale with respect to the
diffusion one.
The bias due to the neglect of diffusion is close to the
1$\sigma$ random uncertainty of the standard models, therefore the systematic
bias due to this source of uncertainty could affect in a significant way the age
estimates obtained using grids which do not take diffusion into account. 

Thus, it would be important to improve the treatment of diffusion in the current 
generation of stellar computations in order to properly follow the evolution of surface chemical abundances.

\section{Comparison with other pipelines}\label{sec:comparison}

The results presented in the previous sections were obtained by using models
from a single
stellar evolutionary code, and they allow a precise characterization of the
impact 
of various input and parameters which influence the stellar evolution.
However, they do not make possible the evaluation of 
systematic biases owing to the adoption of different codes. As it is already
reported in the literature \citep[see e.g.][]{Gai2011, Mathur2012, Chaplin2014},
the adoption of different grids and estimation procedures can produce age
estimates with systematic variance of the same order of the statistical
uncertainties due to the observational errors.

A comparison of our results with those obtained by other techniques will
contribute to assess such a systematic bias. 
In this section we present mass and age estimates of several objects from the
observational samples studied in
\citet{Mathur2012}  and \citet{Chaplin2014} (hereafter M12 and C14
respectively). 

For the first comparison we selected twenty objects analysed
in M12 by adopting RADIUS, YB and SEEK techniques  \citep{Stello2009, Gai2011,
  Quirion2010}. We excluded 
two objects from the original sample, K8760414 and K11713510, because the
observed metallicity of the 
first one is outside our grid and only the SEEK estimates are
available for the second one.  
The seismic and non-seismic observational constraints adopted in the analysis
are presented in Table 2 and 3 of M12; for the solar values we adopt, as in M12, 
$\nu_{\rm max, \sun}$ = 3050 $\mu$Hz and $\Delta \nu_{\sun}$ = 135
$\mu$Hz. Uncertainty of 30 $\mu$Hz on $\nu_{\rm max, \sun}$ and 0.1 $\mu$Hz on
$\Delta \nu_{\sun}$ are considered by error propagation in the uncertainty of
$\nu_{\rm max}$ and  $\Delta \nu$.

\begin{figure*}
\centering
\includegraphics[height=17cm,angle=-90]{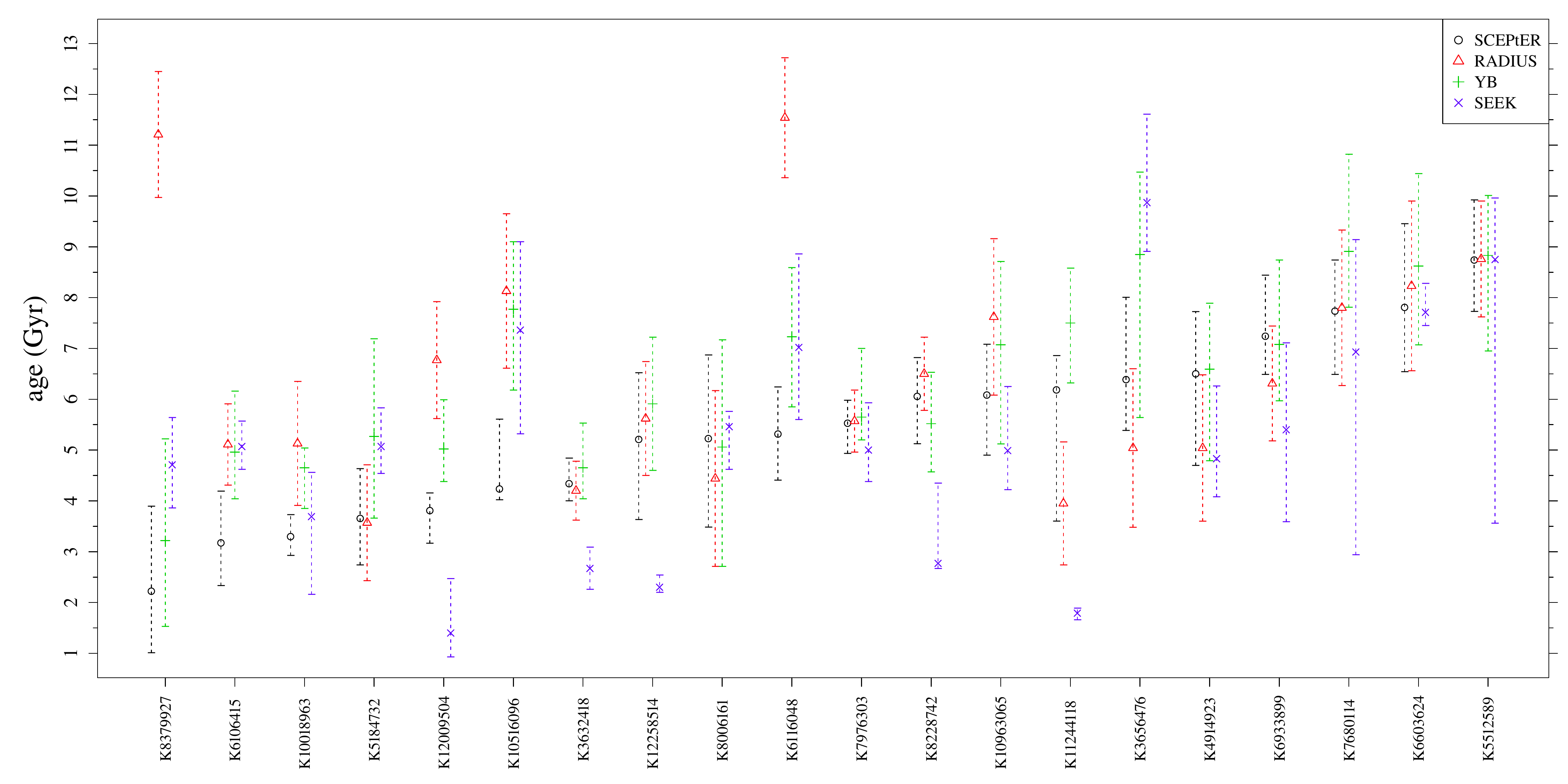}\\
\includegraphics[height=17cm,angle=-90]{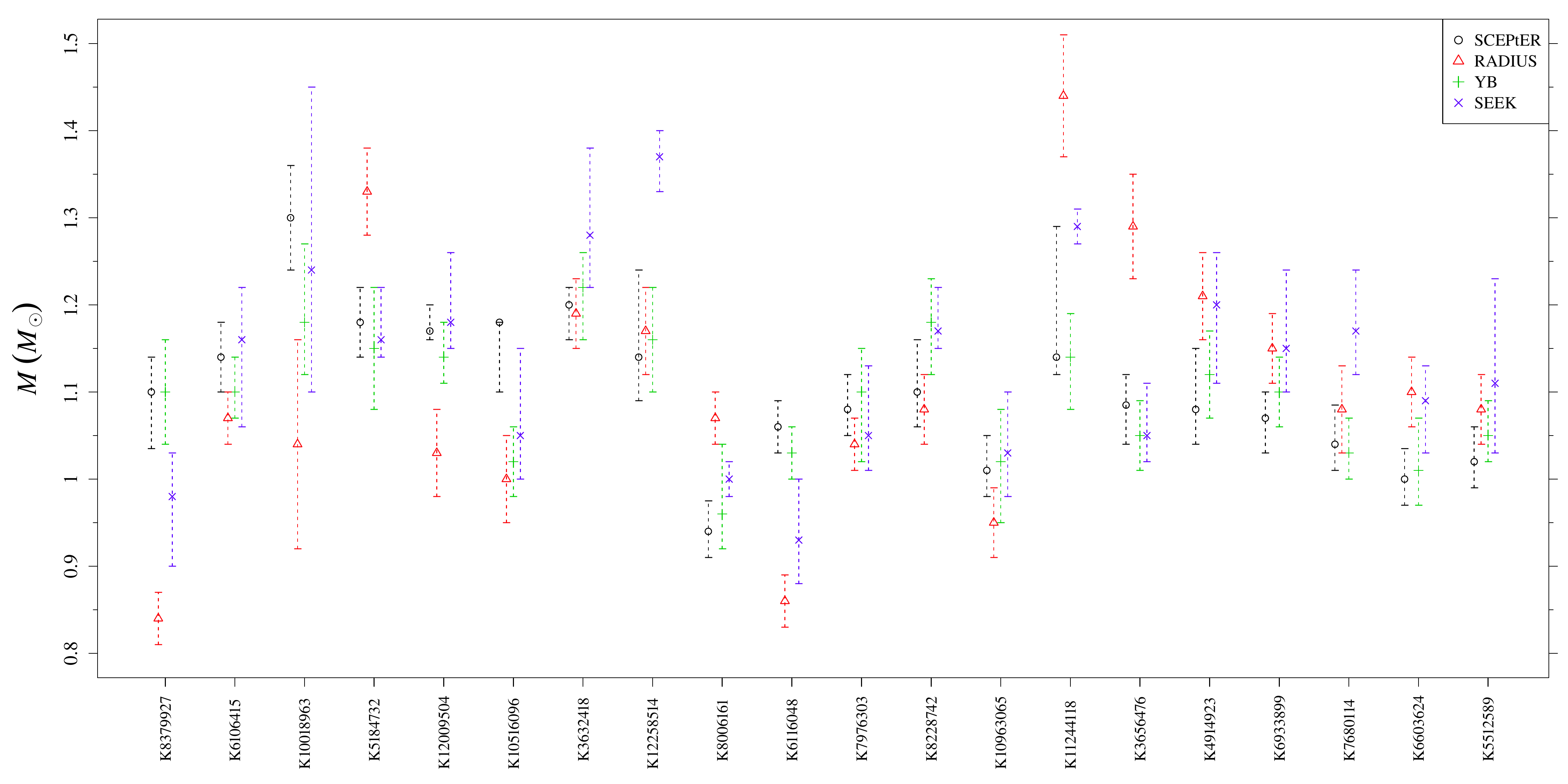}
\caption{SCEPtER age and mass estimates for the observational sample from
  \citet{Mathur2012}, compared with those by RADIUS, YB, and SEEK. Objects are
sorted by ascending SCEPtER estimated age. }
\label{fig:mass-and-age-obs}
\end{figure*}

The SCEPtER  mass and age estimates for the selected sample are in
Table~\ref{tab:estim-obs}. 
Since neither YB nor RADIUS take the evolutionary time scale into account, for
this comparison we adopted unweighted estimates.
In Fig.~\ref{fig:mass-and-age-obs} we compare the
SCEPtER estimates with those of RADIUS, YB, and SEEK reported in M12.
A general agreement among the estimates of the four techniques is
  found. For 10 objects the standard deviation of the pipelines estimates is
  lower than the corresponding random error, obtained by averaging the error
  of the four pipelines; only for 4 objects the computed standard deviation is
greater than two times the random component.
For nine objects (K4914923, K5184732, K5512589,
K6603624, K6933899, K7680114, K7976303, K8006161, and K10963065) age estimates
are all in agreement within their errors. 
Regarding the mass, the same occurs for five objects (K3632418,
K5512589, K6106415, K7976303, and K10963065). Thus age and
mass are in overall agreement for three objects (K5512589, K7976303, and
K10963065).   
However, this comparison is probably not fully appropriate, since the estimates
  of different pipelines are probably highly correlated. To illustrate this
  point let us focus on YB versus SCEPtER comparison, which adopt the same
  estimation scheme. The two techniques  obtain their error bars by scattering 
  the observational values by Monte Carlo perturbation within
  the observational uncertainties. Let us consider an hypothetical comparison
  for which the estimates of the
  two pipelines were performed on the same perturbed set. It can be argued
  that a perturbation that 
  forces YB to overestimate the age with respect to its mean value has a
  similar impact on the SCEPtER estimate. This correlation should be taken into
  account in a comparison of the estimated values, and the usual consistency
  within error bars is probably misleading. 
The actual correlation among estimates could not be evaluated here since we
have not access to the other pipelines to quantify the effect.

A safer comparison involves only the estimates of the different
pipelines disregarding their errors.  
The observational sample is large enough to allow a formal statistical
analysis.  We are interested in possible systematic differences in mass and
age estimates from the four techniques.

The dataset under examination presented, for each star, four estimates
  of mass and four of age. To take into account that for each object repeated
  measurements were 
  available and to increase the statistical power of the test, a
  2-way ANOVA design was adapted to data. Since the four pipeline estimates are
  performed on the same set of stars, the model allows to extract the
  variance due the mean difference among the
  age and mass estimated for the stars in the sample, allowing a more
  powerful assessment of the differences among the various pipelines.
The hypotheses for a parametric ANOVA were not satisfied, due to
a large inhomogeneity of variances in the groups, so the analysis was conducted
adopting the Friedman test, which is the non-parametric analogous of a 2-way
ANOVA \citep{Conover1999}.     

No significant difference was found among the stellar mass estimated by the four
techniques (Friedman $\chi^2$ = 5.89, $df$ = 3, $p$-value = 0.12). This
implies that the hypothesis of no systematic bias due to the choice of a
particular pipeline in the 
recovered mass can not be rejected. However, data provide an indication that
SEEK tends to overestimate the stellar mass with respects to the other
techniques. The median differences of the SEEK masses with respect to those
estimated by SCEPtER, RADIUS, and YB are  0.04, 0.06, and 0.045 $M_{\sun}$
respectively.  
These differences do not reach the significance level due to the low
statistical power attainable with the available sample size.

In the case of age estimates,  a significant difference 
among the pipelines was evidenced (Friedman $\chi^2$ = 16.98, $df$ = 3,
$p$-value = 
0.0071). To assess the origin of this difference we performed a post-hoc Tukey
honest significant differences test \citep[see e.g.][]{hsu1996, snedecor1989},
comparing all the 
possible pairs of mean of age estimates 
between the pipelines. Since the hypotheses of ANOVA are violated, the test is
performed on the rank of the data\footnote{Ranks are obtained by sorting the data into 
ascending order and replacing each value by its relative position in the ordered set.},
 computed for each star \citep{Conover1999}.

The results of the test are presented in Table~\ref{tab:tukey} where we show
the rank differences between groups, the median of the age differences between
groups, and the $p$-values of the tests. 
We found that the SCEPtER age estimates are in agreement with all the
  other
pipelines except YB which gives significantly higher ages with a
median difference of 0.99 Gyr ($p$-value = 0.0069). Regarding the other
pipelines inter-comparisons, SEEK 
age estimates are in median 0.89 Gyr lower than those 
from RADIUS ($p$-value = 0.015) and 0.93 Gyr lower than those from YB
($p$-value = 0.00078). 

As a second comparison we evaluated masses and ages of 73 objects analysed in
C14, 
for which spectroscopic constraints were available from \citet{Bruntt2012}. 
These stars were selected
from the sample of 87 stars reported in Table 2 of C14. We
excluded 14 objects lying outside our estimation grid.

The mass and age estimates presented in C14 have been obtained from  
Bellaterra Stellar Properties Pipeline (BeSPP) coupled 
with a grid constructed
by the GARSTEC code \citep{Weiss2008}; the parameters of the grid
are described 
in \citet{Silva2012}.
The $\Delta \nu$ of each model in this grid was determined
using the calculated frequencies of each model.
The error budget includes the contribution of the systematic
differences due to other pipelines examined in C14.
In this analysis we adopt for the solar values, as in C14, 
$\nu_{\rm max, \sun}$ = 3090 $\mu$Hz and $\Delta \nu_{\sun}$ = 135.1 $\mu$Hz.
Uncertainty of 30 $\mu$Hz on $\nu_{\rm max, \sun}$ and 0.1 $\mu$Hz on
$\Delta \nu_{\sun}$ are considered by error propagation in the uncertainty of
$\nu_{\rm max}$ and  $\Delta \nu$.
In this comparison we adopted weighted estimates, since this effect is
  accounted for in the BeSPP pipeline.

\begin{figure*}
\centering
\includegraphics[height=17cm,angle=-90]{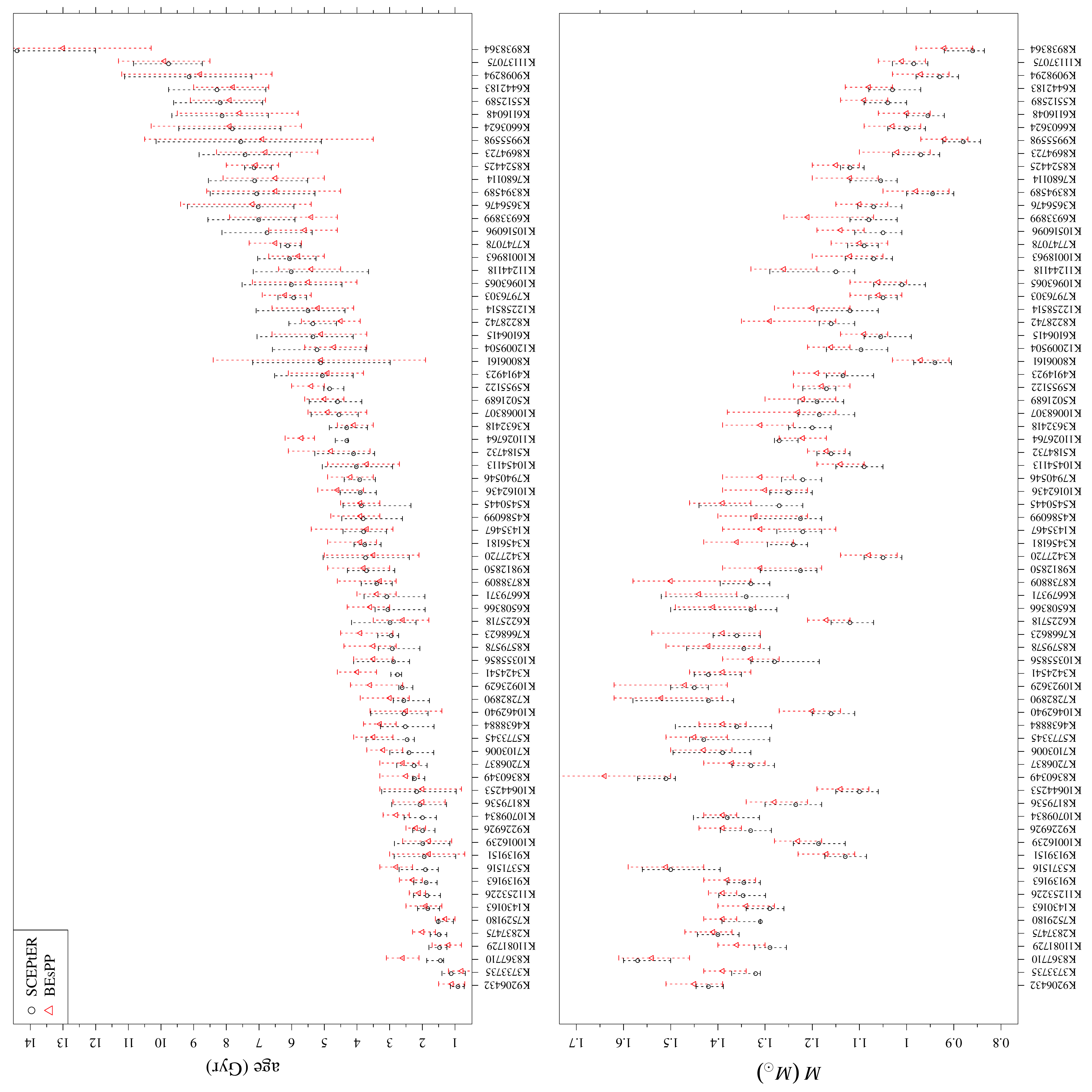}
\caption{SCEPtER weighted age and mass estimates for the observational sample
  from 
  \citet{Chaplin2014}, compared with that by BeSPP.}
\label{fig:mass-and-age-chap}
\end{figure*}

Table~\ref{tab:estim-chap} presents the SCEPtER estimates of age and mass. In
Fig.~\ref{fig:mass-and-age-chap} these estimates are compared with that given
in C14. The estimates are consistent within their errors with the exceptions
of   
K3424541, K8367710, and  K11026764. The same caveat discussed in the
previous comparison about  neglecting 
the estimate correlations applies here.

\begin{table}[ht]
\centering
\caption{SCEPtER unweighted  age and mass estimates for the observational
  sample from 
  \citet{Mathur2012}.} 
\label{tab:estim-obs}
\begin{tabular}{lcc}
  \hline\hline
Star & age (Gyr) & $M$ ($M_{\sun}$)\\
\hline
 K3632418 &  4.34$_{-0.34}^{+0.50}$ &   1.20$_{-0.04}^{+0.02}$\\
 K3656476 &  6.39$_{-1.00}^{+1.62}$ &   1.08$_{-0.05}^{+0.04}$\\
 K4914923 &  6.50$_{-1.80}^{+1.22}$ &   1.08$_{-0.04}^{+0.07}$\\
 K5184732 &  3.65$_{-0.92}^{+0.98}$ &   1.18$_{-0.04}^{+0.04}$\\
 K5512589 &  8.74$_{-1.01}^{+1.18}$ &   1.02$_{-0.03}^{+0.04}$\\
 K6106415 &  3.17$_{-0.84}^{+1.02}$ &   1.14$_{-0.04}^{+0.04}$\\
 K6116048 &  5.32$_{-0.91}^{+0.93}$ &   1.06$_{-0.03}^{+0.03}$\\
 K6603624 &  7.81$_{-1.27}^{+1.65}$ &   1.00$_{-0.03}^{+0.04}$\\
 K6933899 &  7.24$_{-0.75}^{+1.20}$ &   1.07$_{-0.04}^{+0.03}$\\
 K7680114 &  7.73$_{-1.25}^{+1.00}$ &   1.04$_{-0.03}^{+0.04}$\\
 K7976303 &  5.53$_{-0.59}^{+0.45}$ &   1.08$_{-0.03}^{+0.04}$\\
 K8006161 &  5.23$_{-1.74}^{+1.64}$ &   0.94$_{-0.03}^{+0.04}$\\
 K8228742 &  6.06$_{-0.93}^{+0.76}$ &   1.10$_{-0.04}^{+0.06}$\\
 K8379927 &  2.22$_{-1.21}^{+1.67}$ &   1.10$_{-0.06}^{+0.04}$\\
K10018963 &  3.30$_{-0.37}^{+0.43}$ &   1.30$_{-0.06}^{+0.06}$\\
K10516096 &  4.23$_{-0.21}^{+1.37}$ &   1.18$_{-0.08}^{+0.00}$\\
K10963065 &  6.08$_{-1.18}^{+1.00}$ &   1.01$_{-0.03}^{+0.04}$\\
K11244118 &  6.18$_{-2.58}^{+0.67}$ &   1.14$_{-0.02}^{+0.15}$\\
K12009504 &  3.81$_{-0.64}^{+0.35}$ &   1.17$_{-0.01}^{+0.03}$\\
K12258514 &  5.21$_{-1.58}^{+1.31}$ &   1.14$_{-0.05}^{+0.10}$\\
\hline
\end{tabular}
\end{table}

\begin{table}[ht]
\centering
\caption{Age estimations: Tukey multiple comparisons of means for the four
  pipelines examined 
  in the text.}
\label{tab:tukey}
\begin{tabular}{lccc}
  \hline\hline
comparison  & \multicolumn{2}{c}{difference} & $p$-value \\
            & rank       & age (Gyr) &  \\
\hline
RADIUS - SCEPtER &  0.85 &  0.24 & 0.089\\
YB - SCEPtER     &  1.20 &  0.99 & 0.0069\\
SEEK - SCEPtER   & -0.25 & -0.31 & 0.89\\
YB - RADIUS      &  0.35 &  0.18 & 0.76\\
SEEK - RADIUS    & -1.10 & -0.89 & 0.015\\
SEEK - YB        & -1.45 & -0.93 & 0.00078\\
\hline
\end{tabular}
\end{table}

The formal analysis of the differences in the estimates of the two pipelines
was then performed by the paired $t$-test.  
A significant difference ($p$-value < $2 \times 10^{-16}$) was found in the
mass estimates;  
BeSPP estimates of stellar masses were 0.052 $M_{\sun}$ (95\% confidence
interval [0.043 - 0.061] $M_{\sun}$) higher than those of SCEPtER. Beside the
statistical significance, this bias
is also relevant from the stellar evolution point of view, since it is
the same of the average error due to the observational uncertainty,
which is  0.054 $M_{\sun}$ for SCEPtER and  0.068 $M_{\sun}$ for BeSPP.
The difference among age estimates did not reach the
significance 
($p$-value = 0.37); BeSPP age estimates are 0.06 Gyr (95\% confidence
interval [-0.20 - 0.08] Gyr) higher than those of SCEPtER. This bias is
also small with respect to the random errors due to observational uncertainty,
which are about 0.88 Gyr and 1.00 Gyr for SCEPtER and BeSPP respectively, and
it can be safely disregarded from the stellar evolutionary point-of-view. 
These results suggest that the evolutionary time scale is different in the two
stellar grids, since stars of different mass turn out to have a similar age.

\begin{table*}[ht]
\centering
\caption{SCEPtER age and mass estimates for the observational sample from
  \citet{Chaplin2014}.} 
\label{tab:estim-chap}
\begin{tabular}{lcc|lcc|lcc}
  \hline\hline
Star & age (Gyr) & $M$ ($M_{\sun}$) & Star & age (Gyr) & $M$ ($M_{\sun}$) & Star & age (Gyr) & $M$ ($M_{\sun}$)\\
\hline
 K1430163 & 1.83$_{-0.36}^{+0.31}$ & 1.29$_{-0.03}^{+0.05}$ &  K6679371 & 3.09$_{-1.17}^{+0.70}$ & 1.34$_{-0.09}^{+0.18}$ &  K9206432 & 0.91$_{-0.19}^{+0.23}$ & 1.42$_{-0.03}^{+0.03}$ \\
 K1435467 & 3.80$_{-0.70}^{+0.63}$ & 1.22$_{-0.04}^{+0.05}$ &  K6933899 & 7.01$_{-1.11}^{+1.56}$ & 1.08$_{-0.06}^{+0.04}$ &  K9226926 & 1.99$_{-0.38}^{+0.30}$ & 1.33$_{-0.04}^{+0.06}$ \\
 K2837475 & 1.49$_{-0.23}^{+0.27}$ & 1.40$_{-0.04}^{+0.04}$ &  K7103006 & 2.41$_{-0.76}^{+0.59}$ & 1.39$_{-0.06}^{+0.10}$ &  K9812850 & 3.71$_{-0.87}^{+0.58}$ & 1.23$_{-0.04}^{+0.09}$ \\
 K3424541 & 2.76$_{-0.13}^{+0.20}$ & 1.42$_{-0.07}^{+0.03}$ &  K7206837 & 2.25$_{-0.40}^{+0.53}$ & 1.33$_{-0.05}^{+0.04}$ &  K9955598 & 7.56$_{-2.47}^{+2.59}$ & 0.88$_{-0.04}^{+0.04}$ \\
 K3427720 & 3.74$_{-1.34}^{+1.30}$ & 1.05$_{-0.04}^{+0.04}$ &  K7282890 & 2.57$_{-0.79}^{+0.32}$ & 1.42$_{-0.05}^{+0.16}$ & K10016239 & 1.99$_{-0.83}^{+0.87}$ & 1.19$_{-0.06}^{+0.05}$ \\
 K3456181 & 3.76$_{-0.50}^{+0.33}$ & 1.24$_{-0.03}^{+0.05}$ &  K7529180 & 1.51$_{-0.47}^{+0.00}$ & 1.31$_{-0.00}^{+0.08}$ & K10018963 & 6.07$_{-0.81}^{+0.97}$ & 1.07$_{-0.04}^{+0.06}$ \\
 K3632418 & 4.32$_{-0.64}^{+0.53}$ & 1.20$_{-0.04}^{+0.05}$ &  K7668623 & 2.95$_{-0.23}^{+0.41}$ & 1.36$_{-0.05}^{+0.05}$ & K10068307 & 4.55$_{-0.59}^{+0.85}$ & 1.19$_{-0.07}^{+0.04}$ \\
 K3656476 & 7.03$_{-1.09}^{+2.16}$ & 1.07$_{-0.06}^{+0.03}$ &  K7680114 & 7.13$_{-1.62}^{+1.41}$ & 1.05$_{-0.04}^{+0.06}$ & K10162436 & 3.90$_{-0.49}^{+0.61}$ & 1.25$_{-0.05}^{+0.04}$ \\
 K3733735 & 1.12$_{-0.43}^{+0.28}$ & 1.32$_{-0.01}^{+0.05}$ &  K7747078 & 6.12$_{-0.41}^{+0.22}$ & 1.09$_{-0.03}^{+0.03}$ & K10355856 & 2.88$_{-0.49}^{+1.22}$ & 1.28$_{-0.10}^{+0.05}$ \\
 K4586099 & 3.81$_{-1.21}^{+0.65}$ & 1.23$_{-0.05}^{+0.10}$ &  K7940546 & 3.92$_{-0.48}^{+0.47}$ & 1.22$_{-0.04}^{+0.05}$ & K10454113 & 4.02$_{-1.11}^{+1.04}$ & 1.09$_{-0.04}^{+0.06}$ \\
 K4638884 & 2.52$_{-0.88}^{+0.76}$ & 1.36$_{-0.07}^{+0.13}$ &  K7976303 & 5.94$_{-0.39}^{+0.48}$ & 1.05$_{-0.03}^{+0.03}$ & K10462940 & 2.56$_{-0.73}^{+1.02}$ & 1.16$_{-0.05}^{+0.04}$ \\
 K4914923 & 5.06$_{-0.94}^{+1.46}$ & 1.14$_{-0.06}^{+0.03}$ &  K8006161 & 5.11$_{-2.13}^{+2.08}$ & 0.94$_{-0.04}^{+0.04}$ & K10516096 & 6.75$_{-1.38}^{+1.38}$ & 1.05$_{-0.04}^{+0.06}$ \\
 K5021689 & 4.60$_{-0.75}^{+0.86}$ & 1.19$_{-0.06}^{+0.04}$ &  K8179536 & 2.07$_{-0.81}^{+0.86}$ & 1.24$_{-0.05}^{+0.07}$ & K10644253 & 2.17$_{-1.21}^{+1.07}$ & 1.10$_{-0.04}^{+0.05}$ \\
 K5184732 & 4.11$_{-0.65}^{+1.19}$ & 1.16$_{-0.04}^{+0.03}$ &  K8228742 & 5.35$_{-0.72}^{+0.73}$ & 1.16$_{-0.05}^{+0.03}$ & K10709834 & 1.99$_{-0.43}^{+0.56}$ & 1.38$_{-0.07}^{+0.07}$ \\
 K5371516 & 1.90$_{-0.39}^{+0.81}$ & 1.50$_{-0.10}^{+0.06}$ &  K8360349 & 2.24$_{-0.32}^{+0.06}$ & 1.51$_{-0.02}^{+0.06}$ & K10923629 & 2.62$_{-0.33}^{+0.10}$ & 1.45$_{-0.03}^{+0.05}$ \\
 K5450445 & 3.86$_{-1.52}^{+0.56}$ & 1.27$_{-0.05}^{+0.17}$ &  K8367710 & 1.45$_{-0.10}^{+0.42}$ & 1.57$_{-0.07}^{+0.03}$ & K10963065 & 6.01$_{-1.55}^{+1.51}$ & 1.01$_{-0.05}^{+0.06}$ \\
 K5512589 & 8.19$_{-1.30}^{+1.43}$ & 1.04$_{-0.04}^{+0.05}$ &  K8394589 & 7.07$_{-1.78}^{+1.42}$ & 0.94$_{-0.04}^{+0.06}$ & K11026764 & 4.31$_{-0.04}^{+0.35}$ & 1.27$_{-0.04}^{+0.01}$ \\
 K5773345 & 2.47$_{-0.22}^{+1.26}$ & 1.43$_{-0.14}^{+0.03}$ &  K8524425 & 7.15$_{-0.53}^{+0.29}$ & 1.12$_{-0.03}^{+0.02}$ & K11081729 & 1.47$_{-0.23}^{+0.32}$ & 1.29$_{-0.03}^{+0.03}$ \\
 K5955122 & 4.84$_{-0.44}^{+0.18}$ & 1.17$_{-0.02}^{+0.05}$ &  K8579578 & 2.92$_{-0.84}^{+0.43}$ & 1.34$_{-0.06}^{+0.12}$ & K11137075 & 9.77$_{-1.04}^{+1.07}$ & 0.98$_{-0.03}^{+0.04}$ \\
 K6106415 & 5.35$_{-1.23}^{+1.71}$ & 1.05$_{-0.07}^{+0.03}$ &  K8694723 & 7.43$_{-1.39}^{+1.41}$ & 0.97$_{-0.04}^{+0.06}$ & K11244118 & 6.01$_{-2.37}^{+1.16}$ & 1.15$_{-0.04}^{+0.14}$ \\
 K6116048 & 8.13$_{-1.42}^{+1.54}$ & 0.95$_{-0.04}^{+0.04}$ &  K8738809 & 3.40$_{-0.47}^{+0.47}$ & 1.33$_{-0.04}^{+0.06}$ & K11253226 & 1.85$_{-0.41}^{+0.42}$ & 1.35$_{-0.05}^{+0.05}$ \\
 K6225718 & 2.99$_{-0.80}^{+1.18}$ & 1.12$_{-0.05}^{+0.04}$ &  K8938364 &14.41$_{-2.40}^{+1.08}$ & 0.86$_{-0.03}^{+0.06}$ & K12009504 & 5.23$_{-1.50}^{+1.36}$ & 1.10$_{-0.06}^{+0.07}$ \\
 K6442183 & 8.29$_{-1.50}^{+1.48}$ & 1.03$_{-0.06}^{+0.05}$ &  K9098294 & 9.14$_{-1.92}^{+1.98}$ & 0.93$_{-0.04}^{+0.05}$ & K12258514 & 5.50$_{-1.13}^{+1.59}$ & 1.12$_{-0.06}^{+0.07}$ \\
 K6508366 & 3.06$_{-1.14}^{+0.39}$ & 1.33$_{-0.06}^{+0.17}$ &  K9139151 & 1.94$_{-0.96}^{+0.92}$ & 1.13$_{-0.04}^{+0.04}$ &                                &                     \\ 
 K6603624 & 7.82$_{-1.49}^{+1.64}$ & 1.00$_{-0.04}^{+0.04}$ &  K9139163 & 1.88$_{-0.34}^{+0.38}$ & 1.34$_{-0.04}^{+0.04}$ &                                &                     \\
 \hline
\end{tabular}
\end{table*}

The large sample size allowed us a further test on the
  differences between the estimates from SCEPtER and BeSPP
   sub setting the objects into homogeneous groups.
In fact the small difference between the two pipelines may occur either
because they 
provide consistent estimates on the whole set of stars, or because
a compensation between
opposite differences in sub sets of stars occurs.
  
The first step of the analysis was to identify a natural partition of
the dataset, based only on the four observational quantities ($T_{\rm eff}$,
[Fe/H] $\Delta 
\nu$, $\nu_{\rm max}$) adopted in the grids. We performed this step by a
well established statistical procedure that is, an agglomerative 
hierarchical cluster analysis \citep[see e.g.][]{KaufmanRousseeuw90, simar} on
the 73 stellar objects.
Details on the adopted technique are provided in Appendix~\ref{app:cluster}.
Following the analysis, data were split into two groups, the first
containing more massive (interquartile range [1.20 - 1.36] $M_{\sun}$) and
less metallic  
(mean [Fe/H] = $-0.06$ dex) objects with respect to the second one (mean SCEPtER
estimated mass 1.08
$M_{\sun}$ with interquartile range [1.00 - 1.14] $M_{\sun}$, mean [Fe/H] =
0.03 dex).

The subsequent statistical analysis on the two groups (see
e.g. Appendix~\ref{app:cluster}  
for details) revealed 
that the estimates of the two pipelines have an unequal difference in
the two groups. While the median estimates of SCEPtER was
0.19 Gyr higher for less massive objects, for massive stars it was 0.61 Gyr
lower.  
The difference could be due to the fact that the
grid used by BeSPP includes a mild overshooting
and neglects the microscopic diffusion for masses
higher than 1.4 $M_{\sun}$. As seen in Sect.~\ref{sec:errorprop}, both 
the differences lead to higher age estimates for massive models.

\section{Conclusions}\label{sec:conclusions}

We performed a theoretical investigation aimed to quantify the effect of the
current uncertainties in stellar models on the estimates of star ages obtained
by means of grid-based techniques, adopting asteroseismic constraints.
We analysed the uncertainties arising from several input of stellar
models computations, namely input physics, chemical composition, and the
efficiency of microscopic processes.

To this purpose, we used  our grid-based pipeline SCEPtER \citep{scepter1}. We
adopted  
as observational constraints the stellar effective temperature, its metallicity
[Fe/H], the large frequency spacing $\Delta \nu$, and the frequency of maximum
oscillation power $\nu_{\rm max}$ of the star. The grid of stellar models,
computed for the evolutionary phases from ZAMS to the central hydrogen
depletion, has 
been extended to cover 
the mass range [0.8 - 1.6] $M_{\sun}$.

We compared the statistical errors arising
from the uncertainties in observational quantities with the
systematic biases due to the uncertainties in initial helium content, in the
mixing-length parameter value, in the convective core overshooting, and in the
microscopic 
diffusion. We also explored the impact of the uncertainty in radiative
opacities, which is the most relevant input physics with respect to
uncertainty propagation
\citep[see][for a detailed discussion]{incertezze1, incertezze2}. This is the
first time that a comprehensive
detailed theoretical analysis is performed.

We found that the statistical error component in age estimates strongly
depends on the relative
age of the star. The $1 \sigma$ relative error envelope, averaged over all the stellar masses,  is larger than 120\%
and highly asymmetric for stars near the ZAMS,
while it is about 20-30\% and more symmetric at later ages. The
dependence on the stellar 
mass is less important and it is influenced by edge effects. The  $1 \sigma$
envelope, averaged over all the evolutionary stages,  typically  extends from $+42\%$ to $+30\%$ (upper boundary) and from
$-35\%$ to $-20\%$ (lower boundary)
for masses from 0.90 $M_{\sun}$ to 1.40 $M_{\sun}$.
  
We studied the impact of varying the initial helium abundance by changing of
$\pm 1$ the helium-to-metal enrichment ratio $\Delta Y/\Delta Z$.
  The systematic bias is small in the explored range,
except for stars near the ZAMS.  For relative ages older than 0.2 the bias drops
under a value of about 10\%. Overall, the helium bias is about 1/3 of the
width of the envelope due to the random observational uncertainties.

The impact of the uncertainty on radiative opacities was studied here for the
first time.  We found that the current 
uncertainty in radiative opacities -- i.e. $\pm5\%$ -- does not influence the
age estimates from grid techniques. 

The efficiency of the super-adiabatic convection represents one of the weakest
points in theoretical stellar evolution. We studied the impact of varying the  
 mixing-length parameter by computing
 several
synthetic grids with $\alpha_{\rm ml}$ from 1.50 to 1.98, with our solar
calibrated value (i.e. $\alpha_{\rm ml}$ = 1.74) adopted as a reference for
the recovery standard grid. The impact of this source of bias, for models of
mass lower than 1.2 $M_{\sun}$,  was found to be
large with values of about $-20\%$ for $\alpha_{\rm ml}$ = 1.98 and of about
30\% for 
$\alpha_{\rm ml}$ = 1.50. The bias is lower for higher mass models due to the
decreasing thickness of the convective envelope. Therefore the mixing-length
value adopted in the reconstruction can bias the estimates in a significant
way.

The lack of a self-consistent treatment of convection in stars prevents a firm
and 
robust prediction of  convective core extension.
We quantified the resulting bias on grid-based age estimates by 
computing two additional sets of stellar models taking into account convective
core overshooting with
 $\beta$ = 0.2 and 0.4. 
The bias due to the mild-overshooting scenario is at most about -7\% for models
of 1.5 $M_{\sun}$, while for $\beta = 0.4$ it reaches -13\%.
Since these values are small with respect to the standard envelope due to
random errors on the observables adopted in the reconstruction, the
convective core overshooting can be considered as a minor source of bias.

Some grid-based
techniques, such as RADIUS \citep{Stello2009} and SEEK, which both adopt a grid
of 
models computed with the Aarhus STellar Evolution Code \citep{Dalsgaard2008},
do not implement diffusion. We evaluated the bias caused by this neglect and
found that it is very large, reaching values of about
40\% for models of mass lower than 1.1 $M_{\sun}$. 
The
bias is lower for higher mass models due to the fast evolutionary
time scale with respect to the diffusion one. The diffusion induced bias 
is close or greater than the uncertainty arising from random errors in the
observables.

We compared the results obtained by the SCEPtER technique to those found
by adopting other  
grid-based techniques reported in the literature: RADIUS \citep{Stello2009}, YB
\citep{Gai2011}, SEEK \citep{Dalsgaard2008}, and BeSPP \citep{bespp2013}.
We selected a homogeneous 
subset of several targets from the Kepler catalogue, already studied in
\citet{Mathur2012}  and \citet{Chaplin2014}.
No significant difference was found among the stellar
mass estimated by SCEPtER, RADIUS, YB, and SEEK; while age estimates of these
four pipelines significantly differ. Age estimates by YB are significantly
higher than 
SCEPtER ones, with a median difference of 0.99 Gyr; SEEK age
estimates are in median 0.89 Gyr 
lower than those by RADIUS and 0.93 Gyr lower than those by YB.
The comparison of age estimates by SCEPtER and BeSPP showed an overall
agreement, whereas a significant difference was found in the mass estimates
since BeSPP estimation of stellar masses are 0.052 $M_{\sun}$  higher 
than those of SCEPtER. BeSPP age estimates are only 0.07  
Gyr  higher; however we verified that, splitting objects into two
  subgroups, the age of massive objects is generally overestimated by BeSPP.

\begin{acknowledgements}

We thank our anonymous referee for many comments and suggestions that largely
help clarify and improve the paper.
This work has been supported by PRIN-MIUR 2010-2011 ({\em Chemical and dynamical evolution 
of the Milky Way and Local Group galaxies}, PI F. Matteucci), PRIN-INAF 2011 
 ({\em Tracing the formation and evolution of the Galactic Halo with VST}, PI
M. Marconi), and  PRIN-INAF 2012 
 ({\em The M4 Core Project with Hubble Space Telescope}, PI
L. Bedin). 

\end{acknowledgements}

\appendix

\section{Mass and radius estimates}

The statistical error and the systematic bias due to the uncertainty
sources considered in this paper were addressed in V14, but only
for a maximum stellar mass of 1.1 $M_{\sun}$. 
In this appendix we report some results about mass and radius estimate 
obtained in the extended range considered in this paper. 

The results presented in V14 are still valid in this broader range. Since the relative
error on mass and radius is nearly independent on the relative age and on the
mass -- neglecting the edge effect -- we adopt here the same technique as in  
\citet{scepter1} and report for the considered uncertainty source the median
bias of mass and radius relative error and the sample standard deviations. 

\begin{figure*}
\centering
\includegraphics[height=16cm,angle=-90]{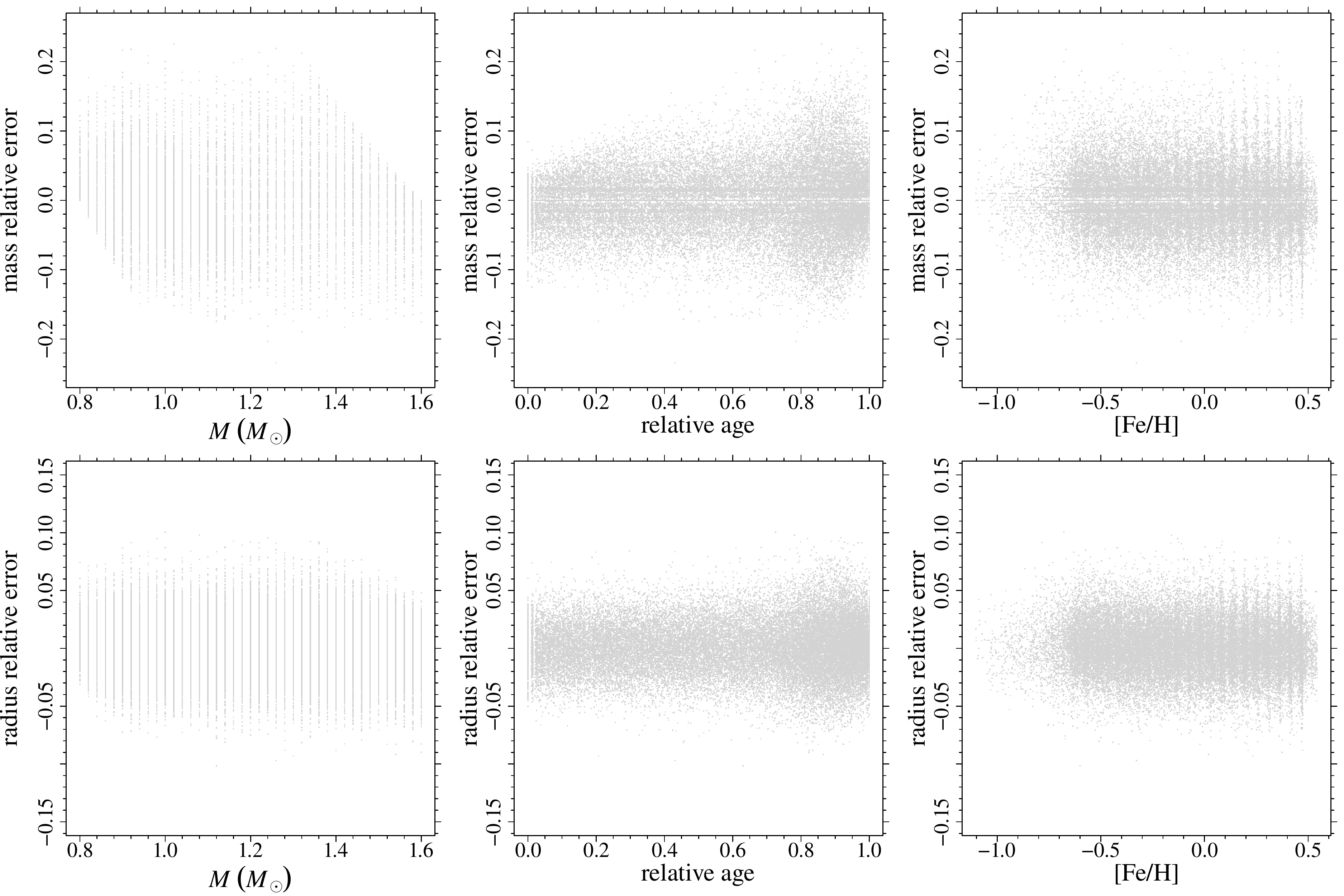}
\caption{{\it Top row}: relative error on mass estimate with respect to the
  true mass of the star, to its relative age and its metallicity [Fe/H]. {\it
    Bottom row}: same as the {\it top row} but for 
  radius relative errors.}
\label{fig:errorMR}
\end{figure*}

Figure~\ref{fig:errorMR} shows, for the standard case of stars sampled and
reconstructed on the standard grid, the trend of mass and radius relative error
versus the true mass of the star, its relative age, and its metallicity.
It is apparent that the same discussion of V14 retains its validity . In
particular it is evident the strong edge effect in the mass panel, and the
inflation of variance at high relative age noted and discussed in the previous
paper of the series.

The analysis of the impact of the considered uncertainties is
summarized in Tab.~\ref{tab:MR}. The biases and 
the standard errors are very similar to the ones reported in
V14. Therefore the results presented in the previous paper remain
valid  also for more massive objects.

\begin{table*}[ht]
\centering
\caption{Summary of mass and radius relative errors.}
\label{tab:MR}
\begin{tabular}{rcccccccc}
  \hline\hline
 & Mass & Std. dev. & $M_{16}$ & $M_{84}$ & Radius & Std. dev. & $R_{16}$ & $R_{84}$ \\ 
  \hline
  standard                &  -0.0005 & 0.0467 & -0.044 & 0.042 & 0.0006 & 0.0228 & -0.022 & 0.023 \\
  weighted                &   0.0003 & 0.0496 & -0.045 & 0.045 & 0.0006 & 0.0234 & -0.022 & 0.023 \\ 
  $\alpha_{\rm ml} = 1.50$  &  -0.0185 & 0.0499 & -0.068 & 0.027 & -0.0068 & 0.0241 & -0.030 & 0.017 \\
  $\alpha_{\rm ml} = 1.98$  &   0.0134 & 0.0484 & -0.030 & 0.058 & 0.0064 & 0.0234 & -0.016 & 0.029 \\ 
  $\Delta Y/\Delta Z = 1$ &  -0.0266 & 0.0505 & -0.075 & 0.019 & -0.0098 & 0.0243 & -0.034 & 0.014 \\
  $\Delta Y/\Delta Z = 3$ &   0.0254 & 0.0497 & -0.019 & 0.071 & 0.0105 & 0.0238 & -0.013 & 0.034 \\ 
  no diffusion            &  -0.0389 & 0.0528 & -0.093 & 0.012 & -0.0146 & 0.0249 & -0.039 & 0.010 \\
  $\beta = 0.2$           &  -0.0040 & 0.0473 & -0.051 & 0.039 & -0.0036 & 0.0243 & -0.027 & 0.020 \\
  $\beta = 0.4$           &   0.0072 & 0.0504 & -0.036 & 0.051 & -0.0004 & 0.0234 & -0.023 & 0.022 \\
   \hline
\end{tabular}
\end{table*}

\section{Differences between SCEPtER and BeSPP age estimates}\label{app:cluster}

The first step of the analysis was to identify a natural partition of
the dataset, based only on the four observational quantities ($T_{\rm eff}$,
[Fe/H] $\Delta 
\nu$, $\nu_{\rm max}$) adopted in the grids. 
 The technique starts with each observation forming a
cluster by itself. 
 Clusters are subsequently merged until only one
cluster 
containing all the observations remains. At each step the two nearest clusters
are 
combined to form one larger cluster. The analysis was performed adopting the
Ward's clustering method to define the cluster similarity
\citep{KaufmanRousseeuw90}.
The computation were performed using R 3.1.0 \citep{R} by mean of the functions
in  the package
{\it cluster} \citep{cluster}.

More technically, let $\mathbf X$ be the matrix of the $p$ observed quantities
for the $n$ objects under consideration. 
Before the analysis the columns of  $\mathbf X$ are standardized that is,
all the columns are scaled to zero mean and unit variance.
Let $x_{ij}$ be the element of the
$i$-th row and $j$-th column of $\mathbf X$. Let $D$ be the $n \times n$
dissimilarity matrix for the $n$ object that is, the matrix whose elements
$d_{ij}$ are the euclidean distances between rows $i$ and $j$ of $\mathbf X$.

\begin{figure*}
\centering
\includegraphics[height=16cm,angle=-90]{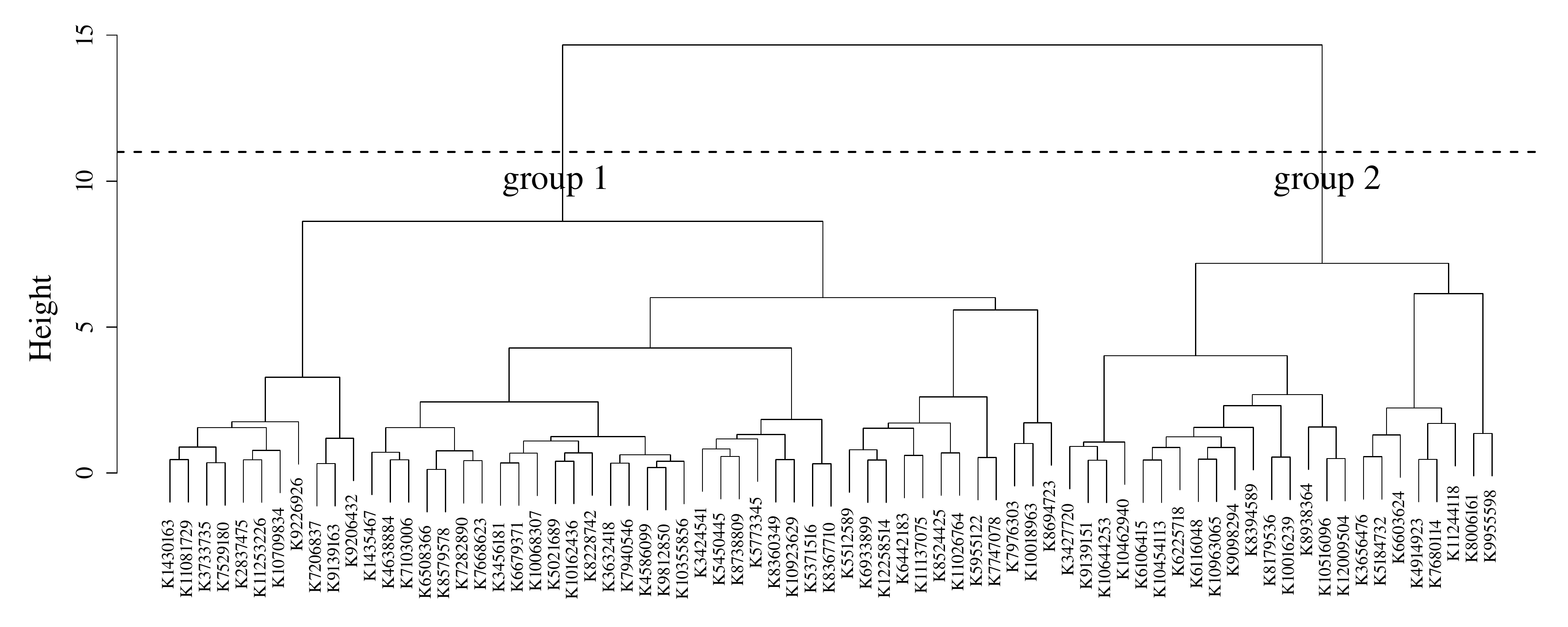}\\
\includegraphics[height=8cm,angle=-90]{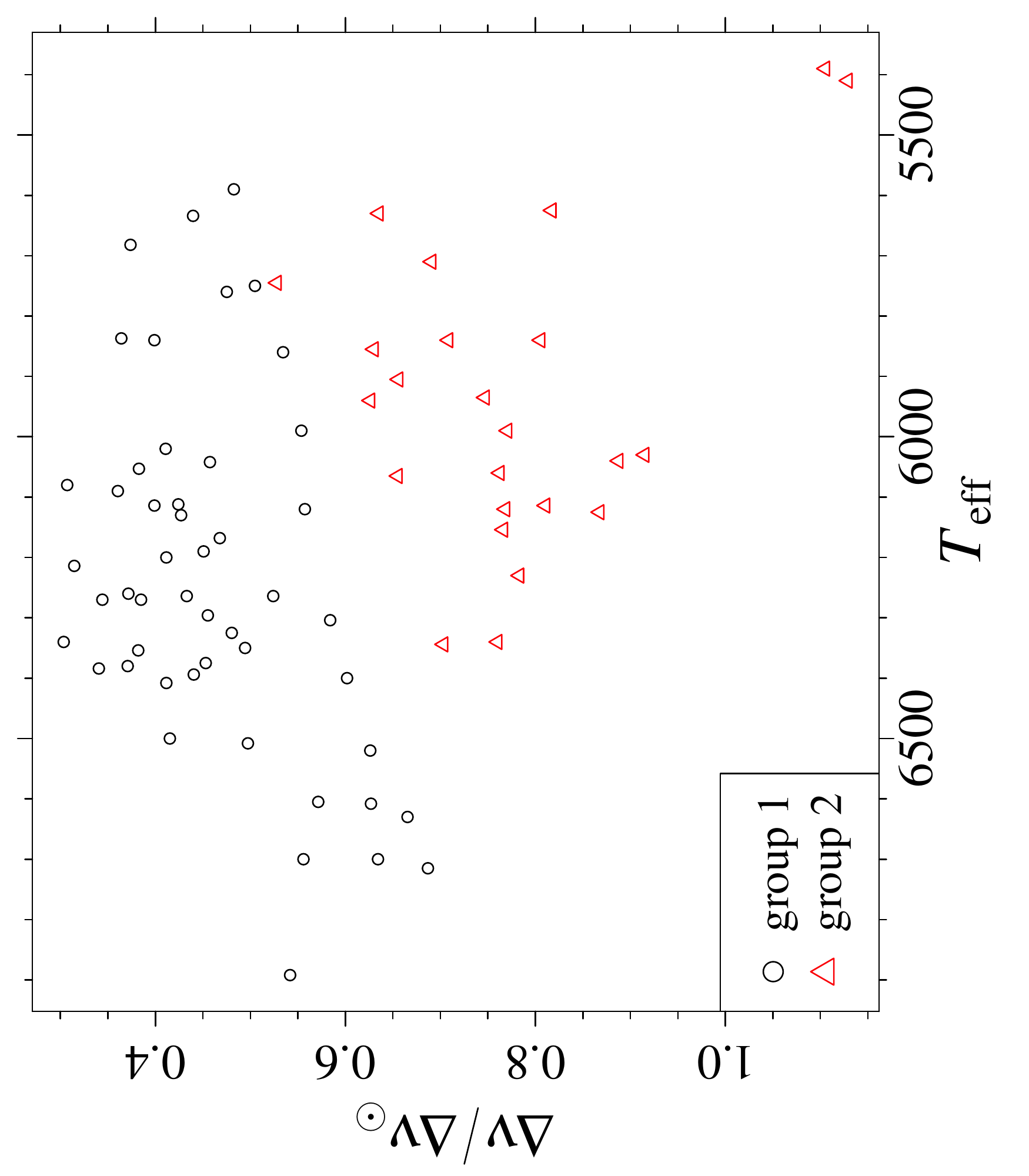}
\includegraphics[height=8cm,angle=-90]{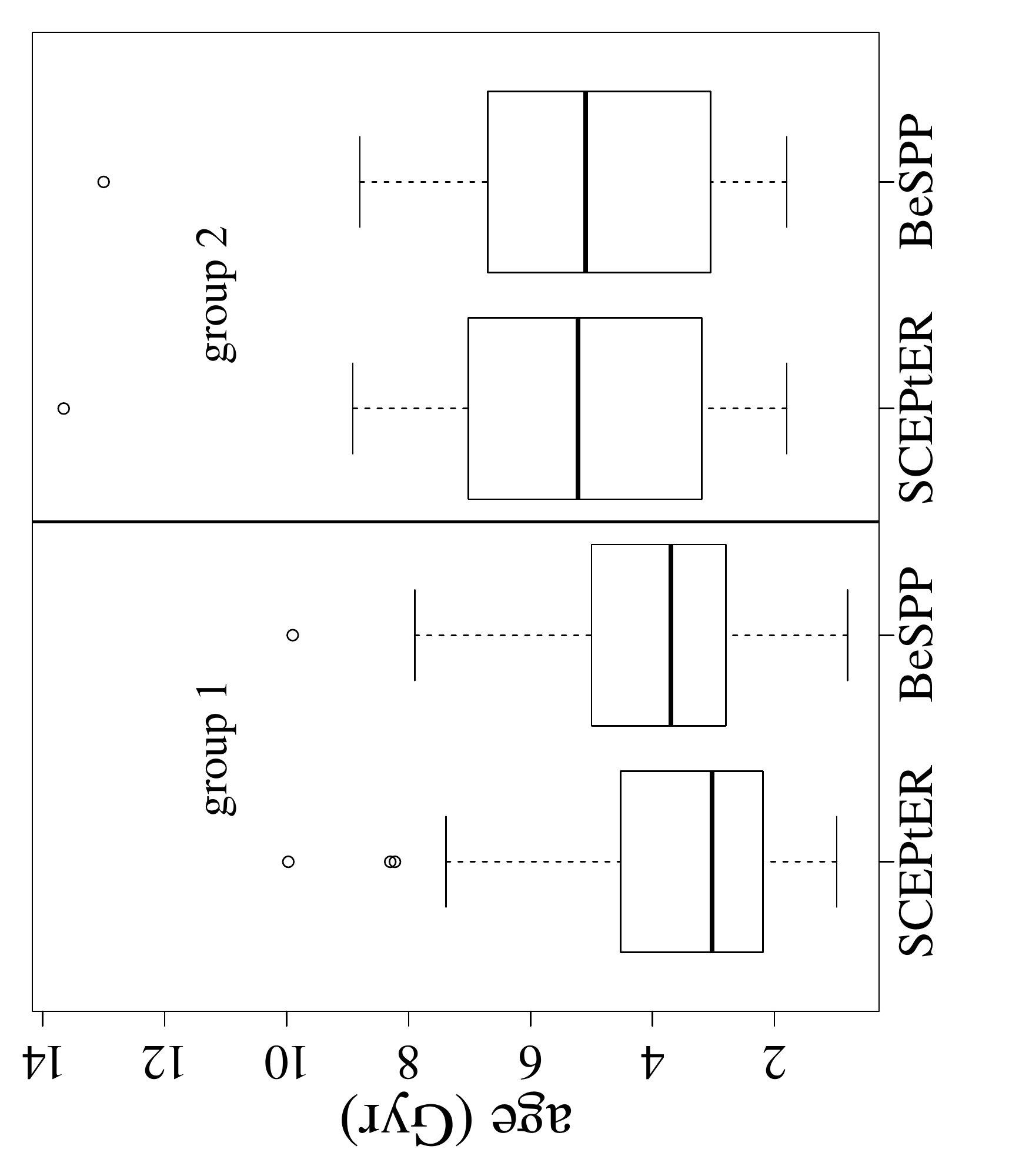}
\caption{{\it Top row}: hierarchical cluster analysis of the observational sample
  adopted in SCEPtER and BeSPP comparison. The dashed line marks the
  sub setting into two groups suggested by the silhouette plot analysis (see
  text). {\it Bottom row, left}: separation of the two group in 
the plane ($T_{\rm eff}$ - $\Delta \nu$). {\it Bottom row, right}: boxplot of age
estimates of the two pipelines in the two groups.}
\label{fig:cluster-ana}
\end{figure*}

At each step, the clustering algorithm  merges the two nearest clusters. The
first step is trivial since each object forms a cluster containing exactly one
object, and therefore the two nearest object are merged. For the second step
a definition of distance among cluster containing more than one object is needed. Let
be $A$ and $B$ two objects joined in a single group $A + B$; the distance
between this group and a group $C$ is
\begin{equation}
d(C, A+B) = \delta_1 d(C,A) + \delta_2 d(C,B) + \delta_3 d(A,B) + \delta_4[d(C,A)
  - d(C,B)] \label{eq:distance}
\end{equation}
Different choices of the weights $\delta_{1,\ldots,4}$ provide different clustering
algorithms. We adopt the Ward's weighting, which minimises the heterogeneity
within clusters \citep[see e.g.][]{simar}
\begin{eqnarray}
\delta_1 & = & \frac{n_C+n_A}{n_A+n_B+n_C}\nonumber\\
\delta_2 & = & \frac{n_C+n_B}{n_A+n_B+n_C}\nonumber\\
\delta_3 & = & -\frac{n_C}{n_A+n_B+n_C}\nonumber\\
\delta_4 & = & 0
\end{eqnarray}
 
The clustering is repeated until all the observations are in the same cluster.
 
The result of the analysis is shown in the dendrogram in the top row of
Fig.~\ref{fig:cluster-ana}. The height of the nodes is the distance, as
defined in Eq.~(\ref{eq:distance}), at which the corresponding clusters merge.
The lower a node, the more similar are the merged clusters.
Cutting the dendrogram at different height (as
done by the dashed line in the figure) 
produces a different number of sub-groups. 
The optimal number of groups suggested by the clustering was determined
according to silhouette plot analysis \citep{Rousseeuw1987,
  KaufmanRousseeuw90} which, for each object, provides a value (silhouette)
which evidences how well the object lies within
its cluster. The  
clustering providing the largest average silhouette is chosen  as the
best one. As a result, the two groups split shown in the figure by the dashed
line turned out to
be the optimal one.

The algorithmic approach to construct a silhouette is the following. For each
object $i$ let be $a(i)$ the average distance among $i$ and all other object
inside the same cluster. Then, for all the other clusters $C$ let $d(i,C)$ be
the average distance of $i$ and all the elements of $C$. Let $b(i)$ be the
minimum value of $d(i,C)$ over all the clusters $C$. The silhouette $s(i)$ is
then
\begin{equation}  
s(i) = \frac{b(i)-a(i)}{{\rm max}(a(i),b(i))}.
\end{equation}  
When a cluster contain a single object then by definition $s(i) = 0$.
An high value of $s(i)$ implies that an observation lie well inside its cluster,
while a value near to 0 indicate that the observation lies equally well
inside it cluster or in the nearest one. A negative silhouette suggests that
the object is possibly in the wrong group.
By averaging the values of $s(i)$ over all the objects one obtains the average
silhouette which is used for diagnostic purposes. 
In our case the average silhouette for a two group clustering is 0.35, while for
a three groups it drops to 0.29. Therefore a 2 group split is preferred.

The group 1 contains more massive (mean SCEPtER estimated mass 1.27
$M_{\sun}$, interquartile range [1.20 - 1.36] $M_{\sun}$) and less metallic 
(mean [Fe/H] = $-0.06$ dex) objects with respect to the second one (mean SCEPtER
estimated mass 1.08
$M_{\sun}$ with interquartile range [1.00 - 1.14] $M_{\sun}$, mean [Fe/H] =
0.03 dex). The separation of the two groups in 
the plane ($T_{\rm eff}$ - $\Delta \nu$) is shown in the bottom row, left
panel of Fig.~\ref{fig:cluster-ana}. The two identified
groups of stars are mainly split by their $\Delta \nu$ values. 

Having defined a grouping for the stars, we repeated the statistical analysis
  of the differences among pipelines, but taking into account also the group
  split.
This was done by adapting a mixed-design model to data \citep[see
  e.g.][]{snedecor1989, 
  linmodR}. This experimental design takes into account the
hierarchy in the data and contains between-objects variables (the group
variable) 
and within-objects variables, allowing to 
account for a different level of variability between and
within objects. In other words, the model includes a level
of variation in addition to the 
per-observation noise term that it is accounted for in common statistical models
such as linear regression models.
The adoption of a nesting variable (the individual stars) allows to take into
account that a couple of
estimate exist for each star.
The other variables in the models are the pipeline adopted for estimations
(categorical variable with level BeSPP and SCEPtER), the subgroup of star
(categorical at two levels), and 
their interaction that is, the variable of interest in the analysis.
A significant interaction means that the two pipeline estimates vary   
differently in the two groups. 
The model was analysed adapting a mixed-design ANOVA
model (also known as repeated measurements ANOVA or split-plot design) to data. 
The result of the analysis is presented in Tab.~\ref{tab:anova}.
The conclusion is that the age estimates by the two pipelines vary
in a significant different way ($p$-value = $2.4 \times 10^{-5}$) from one group
to the other.  
As shown in the right panel in the bottom row of
Fig.~\ref{fig:cluster-ana}, the estimates of the two pipelines are
very close for less massive stars, while those of BeSPP are significantly
higher for massive objects.
The differences in median SCEPtER versus BeSPP age estimates for lighter star
is 0.19 Gyr, while for massive models it is -0.61 Gyr. 
It is relevant to note that the mass estimates does not suffer from a
differential effect like the one present for ages. The same analysis detailed
above, repeated with mass as dependent variable, does not revealed a significant
interaction ($p$-value = 0.40). 

The conclusion of the analysis is therefore that the two pipelines differ in
the evolutionary time scale (the age of 
less massive stars is similar although their mass are estimated different), and
the 
difference in evolutionary time scale changes from massive to light stars.

\begin{table}[ht]
\centering
\caption{ANOVA table for the analysis of the differences between SCEPtER and
  BeSPP. Stars are divided in two sub groups according to the cluster analysis
  (see text).}  
\label{tab:anova}
\begin{tabular}{lrrrrr}
  \hline\hline
 & df & sum sq & mean sq & $F$ & $p$-value \\ 
  \hline
\multicolumn{6}{c}{Error: between stars}\\  
  gruop          & 1  & 73.9   & 73.9 & 7.28 & 0.009 \\ 
  residuals      & 71 & 721.3  & 10.2 &  &  \\ 
 \hline
\multicolumn{6}{c}{Error: within stars}\\
  pipeline       & 1  & 0.139 & 0.139 & 1.02 & 0.317 \\ 
  group:pipeline & 1  & 2.782  & 2.782  & 20.4 & $2.4 \times 10^{-5}$ \\ 
  residuals      & 71 & 9.67  & 0.136 &  &  \\ 
   \hline
\end{tabular}
\tablefoot{Columns legend: df = degree of freedom; sum sq = sum of squares
  (deviance); mean 
sq = mean of squares (variance); $F$ = value of the Fisher's $F$
statistic; $p$-value = significance of the test. The interaction between
group and pipeline is represented by the symbol ``:''. } 
\end{table}

\bibliographystyle{aa}
\bibliography{biblio}

\begin{thebibliography}{64}
\expandafter\ifx\csname natexlab\endcsname\relax\def\natexlab#1{#1}\fi

\bibitem[{{Appourchaux} {et~al.}(2008){Appourchaux}, {Michel}, {Auvergne},
  {Baglin}, {Toutain}, {Baudin}, {Benomar}, {Chaplin}, {Deheuvels}, {Samadi},
  {Verner}, {Boumier}, {Garc{\'{\i}}a}, {Mosser}, {Hulot}, {Ballot}, {Barban},
  {Elsworth}, {Jim{\'e}nez-Reyes}, {Kjeldsen}, {R{\'e}gulo}, \&
  {Roxburgh}}]{Appourchaux2008}
{Appourchaux}, T., {Michel}, E., {Auvergne}, M., {et~al.} 2008, \aap, 488, 705

\bibitem[{{Asplund} {et~al.}(2009){Asplund}, {Grevesse}, {Sauval}, \&
  {Scott}}]{AGSS09}
{Asplund}, M., {Grevesse}, N., {Sauval}, A.~J., \& {Scott}, P. 2009, \araa, 47,
  481

\bibitem[{{Baglin} {et~al.}(2009){Baglin}, {Auvergne}, {Barge}, {Deleuil},
  {Michel}, \& {CoRoT Exoplanet Science Team}}]{Baglin2009}
{Baglin}, A., {Auvergne}, M., {Barge}, P., {et~al.} 2009, in IAU Symposium,
  Vol. 253, IAU Symposium, ed. F.~{Pont}, D.~{Sasselov}, \& M.~J. {Holman},
  71--81

\bibitem[{{Basu} {et~al.}(2010){Basu}, {Chaplin}, \& {Elsworth}}]{Basu2010}
{Basu}, S., {Chaplin}, W.~J., \& {Elsworth}, Y. 2010, \apj, 710, 1596

\bibitem[{{Basu} {et~al.}(2012){Basu}, {Verner}, {Chaplin}, \&
  {Elsworth}}]{Basu2012}
{Basu}, S., {Verner}, G.~A., {Chaplin}, W.~J., \& {Elsworth}, Y. 2012, \apj,
  746, 76

\bibitem[{{Bonaca} {et~al.}(2012){Bonaca}, {Tanner}, {Basu}, {Chaplin},
  {Metcalfe}, {Monteiro}, {Ballot}, {Bedding}, {Bonanno}, {Broomhall},
  {Bruntt}, {Campante}, {Christensen-Dalsgaard}, {Corsaro}, {Elsworth},
  {Garc{\'{\i}}a}, {Hekker}, {Karoff}, {Kjeldsen}, {Mathur}, {R{\'e}gulo},
  {Roxburgh}, {Stello}, {Trampedach}, {Barclay}, {Burke}, \&
  {Caldwell}}]{Bonaca2012}
{Bonaca}, A., {Tanner}, J.~D., {Basu}, S., {et~al.} 2012, \apjl, 755, L12

\bibitem[{{Borucki} {et~al.}(2010){Borucki}, {Koch}, {Basri}, {Batalha},
  {Brown}, {Caldwell}, {Caldwell}, {Christensen-Dalsgaard}, {Cochran},
  {DeVore}, {Dunham}, {Dupree}, {Gautier}, {Geary}, {Gilliland}, {Gould},
  {Howell}, {Jenkins}, {Kondo}, {Latham}, {Marcy}, {Meibom}, {Kjeldsen},
  {Lissauer}, {Monet}, {Morrison}, {Sasselov}, {Tarter}, {Boss}, {Brownlee},
  {Owen}, {Buzasi}, {Charbonneau}, {Doyle}, {Fortney}, {Ford}, {Holman},
  {Seager}, {Steffen}, {Welsh}, {Rowe}, {Anderson}, {Buchhave}, {Ciardi},
  {Walkowicz}, {Sherry}, {Horch}, {Isaacson}, {Everett}, {Fischer}, {Torres},
  {Johnson}, {Endl}, {MacQueen}, {Bryson}, {Dotson}, {Haas}, {Kolodziejczak},
  {Van Cleve}, {Chandrasekaran}, {Twicken}, {Quintana}, {Clarke}, {Allen},
  {Li}, {Wu}, {Tenenbaum}, {Verner}, {Bruhweiler}, {Barnes}, \&
  {Prsa}}]{Borucki2010}
{Borucki}, W.~J., {Koch}, D., {Basri}, G., {et~al.} 2010, Science, 327, 977

\bibitem[{{Bruntt} {et~al.}(2012){Bruntt}, {Basu}, {Smalley}, {Chaplin},
  {Verner}, {Bedding}, {Catala}, {Gazzano}, {Molenda-{\.Z}akowicz}, {Thygesen},
  {Uytterhoeven}, {Hekker}, {Huber}, {Karoff}, {Mathur}, {Mosser},
  {Appourchaux}, {Campante}, {Elsworth}, {Garc{\'{\i}}a}, {Handberg},
  {Metcalfe}, {Quirion}, {R{\'e}gulo}, {Roxburgh}, {Stello},
  {Christensen-Dalsgaard}, {Kawaler}, {Kjeldsen}, {Morris}, {Quintana}, \&
  {Sanderfer}}]{Bruntt2012}
{Bruntt}, H., {Basu}, S., {Smalley}, B., {et~al.} 2012, \mnras, 423, 122

\bibitem[{{Casagrande} {et~al.}(2011){Casagrande}, {Sch{\"o}nrich}, {Asplund},
  {Cassisi}, {Ram{\'{\i}}rez}, {Mel{\'e}ndez}, {Bensby}, \&
  {Feltzing}}]{Casagrande2011}
{Casagrande}, L., {Sch{\"o}nrich}, R., {Asplund}, M., {et~al.} 2011, \aap, 530,
  A138

\bibitem[{{Chaboyer} {et~al.}(2001){Chaboyer}, {Fenton}, {Nelan}, {Patnaude},
  \& {Simon}}]{Chaboyer2001}
{Chaboyer}, B., {Fenton}, W.~H., {Nelan}, J.~E., {Patnaude}, D.~J., \& {Simon},
  F.~E. 2001, \apj, 562, 521

\bibitem[{{Chaplin} {et~al.}(2014){Chaplin}, {Basu}, {Huber}, {Serenelli},
  {Casagrande}, {Silva Aguirre}, {Ball}, {Creevey}, {Gizon}, {Handberg},
  {Karoff}, {Lutz}, {Marques}, {Miglio}, {Stello}, {Suran}, {Pricopi},
  {Metcalfe}, {Monteiro}, {Molenda-{\.Z}akowicz}, {Appourchaux},
  {Christensen-Dalsgaard}, {Elsworth}, {Garc{\'{\i}}a}, {Houdek}, {Kjeldsen},
  {Bonanno}, {Campante}, {Corsaro}, {Gaulme}, {Hekker}, {Mathur}, {Mosser},
  {R{\'e}gulo}, \& {Salabert}}]{Chaplin2014}
{Chaplin}, W.~J., {Basu}, S., {Huber}, D., {et~al.} 2014, \apjs, 210, 1

\bibitem[{{Christensen-Dalsgaard}(2008)}]{Dalsgaard2008}
{Christensen-Dalsgaard}, J. 2008, \apss, 316, 13

\bibitem[{{Clausen} {et~al.}(2009){Clausen}, {Bruntt}, {Claret}, {Larsen},
  {Andersen}, {Nordstr{\"o}m}, \& {Gim{\'e}nez}}]{Clausen2009}
{Clausen}, J.~V., {Bruntt}, H., {Claret}, A., {et~al.} 2009, \aap, 502, 253

\bibitem[{Conover(1999)}]{Conover1999}
Conover, W. 1999, Practical nonparametric statistics, Wiley series in
  probability and statistics: Applied probability and statistics (Wiley)

\bibitem[{{Cyburt} {et~al.}(2004){Cyburt}, {Fields}, \& {Olive}}]{cyburt04}
{Cyburt}, R.~H., {Fields}, B.~D., \& {Olive}, K.~A. 2004, \prd, 69, 123519

\bibitem[{{Degl'Innocenti} {et~al.}(2008){Degl'Innocenti}, {Prada Moroni},
  {Marconi}, \& {Ruoppo}}]{scilla2008}
{Degl'Innocenti}, S., {Prada Moroni}, P.~G., {Marconi}, M., \& {Ruoppo}, A.
  2008, \apss, 316, 25

\bibitem[{{Deheuvels} \& {Michel}(2011)}]{Deheuvels2011}
{Deheuvels}, S. \& {Michel}, E. 2011, \aap, 535, A91

\bibitem[{Dell'Omodarme \& Valle(2013)}]{stellar}
Dell'Omodarme, M. \& Valle, G. 2013, The R Journal, 5, 108

\bibitem[{{Dell'Omodarme} {et~al.}(2012){Dell'Omodarme}, {Valle},
  {Degl'Innocenti}, \& {Prada Moroni}}]{database2012}
{Dell'Omodarme}, M., {Valle}, G., {Degl'Innocenti}, S., \& {Prada Moroni},
  P.~G. 2012, A\&A, 540, A26

\bibitem[{{Epstein} \& {Pinsonneault}(2014)}]{Epstein2014}
{Epstein}, C.~R. \& {Pinsonneault}, M.~H. 2014, \apj, 780, 159

\bibitem[{{Faraway}(2004)}]{linmodR}
{Faraway}, J.~J. 2004, Linear Models with R (Chapman \& Hall/CRC)

\bibitem[{{Gai} {et~al.}(2011){Gai}, {Basu}, {Chaplin}, \&
  {Elsworth}}]{Gai2011}
{Gai}, N., {Basu}, S., {Chaplin}, W.~J., \& {Elsworth}, Y. 2011, \apj, 730, 63

\bibitem[{{Gennaro} {et~al.}(2010){Gennaro}, {Prada Moroni}, \&
  {Degl'Innocenti}}]{gennaro10}
{Gennaro}, M., {Prada Moroni}, P.~G., \& {Degl'Innocenti}, S. 2010, \aap, 518,
  A13+

\bibitem[{{Gilliland} {et~al.}(2010){Gilliland}, {Brown},
  {Christensen-Dalsgaard}, {Kjeldsen}, {Aerts}, {Appourchaux}, {Basu},
  {Bedding}, {Chaplin}, {Cunha}, {De Cat}, {De Ridder}, {Guzik}, {Handler},
  {Kawaler}, {Kiss}, {Kolenberg}, {Kurtz}, {Metcalfe}, {Monteiro}, {Szab{\'o}},
  {Arentoft}, {Balona}, {Debosscher}, {Elsworth}, {Quirion}, {Stello},
  {Su{\'a}rez}, {Borucki}, {Jenkins}, {Koch}, {Kondo}, {Latham}, {Rowe}, \&
  {Steffen}}]{Gilliland2010}
{Gilliland}, R.~L., {Brown}, T.~M., {Christensen-Dalsgaard}, J., {et~al.} 2010,
  \pasp, 122, 131

\bibitem[{H{\"a}rdle \& Simar(2012)}]{simar}
H{\"a}rdle, W.~K. \& Simar, L. 2012, Applied Multivariate Statistical Analysis
  (Springer)

\bibitem[{Hsu(1996)}]{hsu1996}
Hsu, J. 1996, Multiple Comparisons: Theory and Methods (Taylor \& Francis)

\bibitem[{{Huber} {et~al.}(2013){Huber}, {Chaplin}, {Christensen-Dalsgaard},
  {Gilliland}, {Kjeldsen}, {Buchhave}, {Fischer}, {Lissauer}, {Rowe},
  {Sanchis-Ojeda}, {Basu}, {Handberg}, {Hekker}, {Howard}, {Isaacson},
  {Karoff}, {Latham}, {Lund}, {Lundkvist}, {Marcy}, {Miglio}, {Silva Aguirre},
  {Stello}, {Arentoft}, {Barclay}, {Bedding}, {Burke}, {Christiansen},
  {Elsworth}, {Haas}, {Kawaler}, {Metcalfe}, {Mullally}, \&
  {Thompson}}]{Huber2013}
{Huber}, D., {Chaplin}, W.~J., {Christensen-Dalsgaard}, J., {et~al.} 2013,
  \apj, 767, 127

\bibitem[{{Jimenez} {et~al.}(2003){Jimenez}, {Flynn}, {MacDonald}, \&
  {Gibson}}]{jimenez03}
{Jimenez}, R., {Flynn}, C., {MacDonald}, J., \& {Gibson}, B.~K. 2003, Science,
  299, 1552

\bibitem[{{J{\o}rgensen} \& {Lindegren}(2005)}]{Jorgensen2005}
{J{\o}rgensen}, B.~R. \& {Lindegren}, L. 2005, \aap, 436, 127

\bibitem[{{Kaufman} \& {Rousseeuw}(1990)}]{KaufmanRousseeuw90}
{Kaufman}, L. \& {Rousseeuw}, P.~J. 1990, Finding groups in data: an
  introduction to cluster analysis (New York: John Wiley and Sons)

\bibitem[{{Kjeldsen} \& {Bedding}(1995)}]{Kjeldsen1995}
{Kjeldsen}, H. \& {Bedding}, T.~R. 1995, \aap, 293, 87

\bibitem[{{Lebreton}(2013)}]{Lebreton2013}
{Lebreton}, Y. 2013, in EAS Publications Series, Vol.~63, EAS Publications
  Series, 123--133

\bibitem[{{Lebreton} \& {Montalb{\'a}n}(2009)}]{Lebreton2009}
{Lebreton}, Y. \& {Montalb{\'a}n}, J. 2009, in IAU Symposium, Vol. 258, IAU
  Symposium, ed. E.~E. {Mamajek}, D.~R. {Soderblom}, \& R.~F.~G. {Wyse},
  419--430

\bibitem[{{Maechler} {et~al.}(2014){Maechler}, {Rousseeuw}, {Struyf}, {Hubert},
  \& {Hornik}}]{cluster}
{Maechler}, M., {Rousseeuw}, P., {Struyf}, A., {Hubert}, M., \& {Hornik}, K.
  2014, cluster: Cluster Analysis Basics and Extensions, r package version
  1.15.2 --- For new features, see the 'Changelog' file (in the package source)

\bibitem[{{Magic} {et~al.}(2014){Magic}, {Weiss}, \& {Asplund}}]{Magic2014}
{Magic}, Z., {Weiss}, A., \& {Asplund}, M. 2014, ArXiv e-prints

\bibitem[{{Mathur} {et~al.}(2012){Mathur}, {Metcalfe}, {Woitaszek}, {Bruntt},
  {Verner}, {Christensen-Dalsgaard}, {Creevey}, {Do{\v g}an}, {Basu}, {Karoff},
  {Stello}, {Appourchaux}, {Campante}, {Chaplin}, {Garc{\'{\i}}a}, {Bedding},
  {Benomar}, {Bonanno}, {Deheuvels}, {Elsworth}, {Gaulme}, {Guzik}, {Handberg},
  {Hekker}, {Herzberg}, {Monteiro}, {Piau}, {Quirion}, {R{\'e}gulo}, {Roth},
  {Salabert}, {Serenelli}, {Thompson}, {Trampedach}, {White}, {Ballot},
  {Brand{\~a}o}, {Molenda-{\.Z}akowicz}, {Kjeldsen}, {Twicken}, {Uddin}, \&
  {Wohler}}]{Mathur2012}
{Mathur}, S., {Metcalfe}, T.~S., {Woitaszek}, M., {et~al.} 2012, \apj, 749, 152

\bibitem[{{Metcalfe} {et~al.}(2014){Metcalfe}, {Creevey}, {Dogan}, {Mathur},
  {Xu}, {Bedding}, {Chaplin}, {Christensen-Dalsgaard}, {Karoff}, {Trampedach},
  {Benomar}, {Brown}, {Buzasi}, {Campante}, {Celik}, {Cunha}, {Davies},
  {Deheuvels}, {Derekas}, {Di Mauro}, {Garcia}, {Guzik}, {Howe}, {MacGregor},
  {Mazumdar}, {Montalban}, {Monteiro}, {Salabert}, {Serenelli}, {Stello},
  {Steslicki}, {Suran}, {Yildiz}, {Aksoy}, {Elsworth}, {Gruberbauer},
  {Guenther}, {Lebreton}, {Molaverdikhani}, {Pricopi}, {Simoniello}, \&
  {White}}]{Metcalfe2014}
{Metcalfe}, T.~S., {Creevey}, O.~L., {Dogan}, G., {et~al.} 2014, ArXiv e-prints

\bibitem[{{Michel} {et~al.}(2008){Michel}, {Baglin}, {Auvergne}, {Catala},
  {Samadi}, {Baudin}, {Appourchaux}, {Barban}, {Weiss}, {Berthomieu},
  {Boumier}, {Dupret}, {Garcia}, {Fridlund}, {Garrido}, {Goupil}, {Kjeldsen},
  {Lebreton}, {Mosser}, {Grotsch-Noels}, {Janot-Pacheco}, {Provost},
  {Roxburgh}, {Thoul}, {Toutain}, {Tiph{\`e}ne}, {Turck-Chieze}, {Vauclair},
  {Vauclair}, {Aerts}, {Alecian}, {Ballot}, {Charpinet}, {Hubert},
  {Ligni{\`e}res}, {Mathias}, {Monteiro}, {Neiner}, {Poretti}, {Renan de
  Medeiros}, {Ribas}, {Rieutord}, {Cort{\'e}s}, \& {Zwintz}}]{Michel2008}
{Michel}, E., {Baglin}, A., {Auvergne}, M., {et~al.} 2008, Science, 322, 558

\bibitem[{{Pagel} \& {Portinari}(1998)}]{pagel98}
{Pagel}, B.~E.~J. \& {Portinari}, L. 1998, \mnras, 298, 747

\bibitem[{{Peimbert} {et~al.}(2007{\natexlab{a}}){Peimbert}, {Luridiana}, \&
  {Peimbert}}]{peimbert07a}
{Peimbert}, M., {Luridiana}, V., \& {Peimbert}, A. 2007{\natexlab{a}}, \apj,
  666, 636

\bibitem[{{Peimbert} {et~al.}(2007{\natexlab{b}}){Peimbert}, {Luridiana},
  {Peimbert}, \& {Carigi}}]{peimbert07b}
{Peimbert}, M., {Luridiana}, V., {Peimbert}, A., \& {Carigi}, L.
  2007{\natexlab{b}}, in Astronomical Society of the Pacific Conference Series,
  Vol. 374, From Stars to Galaxies: Building the Pieces to Build Up the
  Universe, ed. {A.~Vallenari, R.~Tantalo, L.~Portinari, \& A.~Moretti}, 81--+

\bibitem[{{Pont} \& {Eyer}(2004)}]{Pont2004}
{Pont}, F. \& {Eyer}, L. 2004, \mnras, 351, 487

\bibitem[{{Quirion} {et~al.}(2010){Quirion}, {Christensen-Dalsgaard}, \&
  {Arentoft}}]{Quirion2010}
{Quirion}, P.-O., {Christensen-Dalsgaard}, J., \& {Arentoft}, T. 2010, \apj,
  725, 2176

\bibitem[{{R Development Core Team}(2014)}]{R}
{R Development Core Team}. 2014, R: A Language and Environment for Statistical
  Computing, R Foundation for Statistical Computing, Vienna, Austria

\bibitem[{{Rousseeuw}(1987)}]{Rousseeuw1987}
{Rousseeuw}, P.~J. 1987, Journal of Computational and Applied Mathematics, 20,
  53

\bibitem[{{Scott}(1992)}]{Scott1992}
{Scott}, D.~W. 1992, Multivariate Density Estimation. Theory, Practice and
  Visualization (Wiley)

\bibitem[{{Serenelli} {et~al.}(2013){Serenelli}, {Bergemann}, {Ruchti}, \&
  {Casagrande}}]{bespp2013}
{Serenelli}, A.~M., {Bergemann}, M., {Ruchti}, G., \& {Casagrande}, L. 2013,
  \mnras, 429, 3645

\bibitem[{{Silva Aguirre} {et~al.}(2013){Silva Aguirre}, {Basu}, {Brand{\~a}o},
  {Christensen-Dalsgaard}, {Deheuvels}, {Do{\u g}an}, {Metcalfe}, {Serenelli},
  {Ballot}, {Chaplin}, {Cunha}, {Weiss}, {Appourchaux}, {Casagrande},
  {Cassisi}, {Creevey}, {Garc{\'{\i}}a}, {Lebreton}, {Noels}, {Sousa},
  {Stello}, {White}, {Kawaler}, \& {Kjeldsen}}]{SilvaAguirre2013}
{Silva Aguirre}, V., {Basu}, S., {Brand{\~a}o}, I.~M., {et~al.} 2013, \apj,
  769, 141

\bibitem[{{Silva Aguirre} {et~al.}(2012){Silva Aguirre}, {Casagrande}, {Basu},
  {Campante}, {Chaplin}, {Huber}, {Miglio}, {Serenelli}, {Ballot}, {Bedding},
  {Christensen-Dalsgaard}, {Creevey}, {Elsworth}, {Garc{\'{\i}}a}, {Gilliland},
  {Hekker}, {Kjeldsen}, {Mathur}, {Metcalfe}, {Monteiro}, {Mosser},
  {Pinsonneault}, {Stello}, {Weiss}, {Tenenbaum}, {Twicken}, \&
  {Uddin}}]{Silva2012}
{Silva Aguirre}, V., {Casagrande}, L., {Basu}, S., {et~al.} 2012, \apj, 757, 99

\bibitem[{Snedecor \& Cochran(1989)}]{snedecor1989}
Snedecor, G. \& Cochran, W. 1989, Statistical methods, Statistical Methods No.
  v. 276 (Iowa State University Press)

\bibitem[{{Soderblom}(2010)}]{Soderblom2010}
{Soderblom}, D.~R. 2010, \araa, 48, 581

\bibitem[{{Steigman}(2006)}]{steigman06}
{Steigman}, G. 2006, International Journal of Modern Physics E, 15, 1

\bibitem[{{Stello} {et~al.}(2009){Stello}, {Chaplin}, {Bruntt}, {Creevey},
  {Garc{\'{\i}}a-Hern{\'a}ndez}, {Monteiro}, {Moya}, {Quirion}, {Sousa},
  {Su{\'a}rez}, {Appourchaux}, {Arentoft}, {Ballot}, {Bedding},
  {Christensen-Dalsgaard}, {Elsworth}, {Fletcher}, {Garc{\'{\i}}a}, {Houdek},
  {Jim{\'e}nez-Reyes}, {Kjeldsen}, {New}, {R{\'e}gulo}, {Salabert}, \&
  {Toutain}}]{Stello2009}
{Stello}, D., {Chaplin}, W.~J., {Bruntt}, H., {et~al.} 2009, \apj, 700, 1589

\bibitem[{{Tanner} {et~al.}(2014){Tanner}, {Basu}, \& {Demarque}}]{Tanner2014}
{Tanner}, J.~D., {Basu}, S., \& {Demarque}, P. 2014, \apjl, 785, L13

\bibitem[{{Thoul} {et~al.}(1994){Thoul}, {Bahcall}, \& {Loeb}}]{thoul94}
{Thoul}, A.~A., {Bahcall}, J.~N., \& {Loeb}, A. 1994, \apj, 421, 828

\bibitem[{{Trampedach} \& {Stein}(2011)}]{Trampedach2011}
{Trampedach}, R. \& {Stein}, R.~F. 2011, \apj, 731, 78

\bibitem[{{Ulrich}(1986)}]{Ulrich1986}
{Ulrich}, R.~K. 1986, \apjl, 306, L37

\bibitem[{{Valle} {et~al.}(2013{\natexlab{a}}){Valle}, {Dell'Omodarme}, {Prada
  Moroni}, \& {Degl'Innocenti}}]{incertezze1}
{Valle}, G., {Dell'Omodarme}, M., {Prada Moroni}, P.~G., \& {Degl'Innocenti},
  S. 2013{\natexlab{a}}, \aap, 549, A50

\bibitem[{{Valle} {et~al.}(2013{\natexlab{b}}){Valle}, {Dell'Omodarme}, {Prada
  Moroni}, \& {Degl'Innocenti}}]{incertezze2}
{Valle}, G., {Dell'Omodarme}, M., {Prada Moroni}, P.~G., \& {Degl'Innocenti},
  S. 2013{\natexlab{b}}, \aap, 554, A68

\bibitem[{{Valle} {et~al.}(2014){Valle}, {Dell'Omodarme}, {Prada Moroni}, \&
  {Degl'Innocenti}}]{scepter1}
{Valle}, G., {Dell'Omodarme}, M., {Prada Moroni}, P.~G., \& {Degl'Innocenti},
  S. 2014, \aap, 561, A125 (V14)

\bibitem[{{Valle} {et~al.}(2009){Valle}, {Marconi}, {Degl'Innocenti}, \& {Prada
  Moroni}}]{cefeidi}
{Valle}, G., {Marconi}, M., {Degl'Innocenti}, S., \& {Prada Moroni}, P.~G.
  2009, \aap, 507, 1541

\bibitem[{Venables \& Ripley(2002)}]{venables2002modern}
Venables, W. \& Ripley, B. 2002, Modern applied statistics with S, Statistics
  and computing (Springer)

\bibitem[{{Weiss} \& {Schlattl}(2008)}]{Weiss2008}
{Weiss}, A. \& {Schlattl}, H. 2008, \apss, 316, 99

\bibitem[{{Y{\i}ld{\i}z}(2007)}]{Yildiz2007}
{Y{\i}ld{\i}z}, M. 2007, \mnras, 374, 1264

\end{thebibliography}

\end{document}